\documentclass[vecphys]{svmult}

\usepackage{makeidx}         % allows index generation
\usepackage{graphicx}        % standard LaTeX graphics tool
\usepackage{psfig}           % alternative graphics tool
\usepackage{multicol}        % used for the two-column index
\usepackage{url}             % allows inclusion of URLs
\usepackage[bottom]{footmisc}% places footnotes at page bottom

\newcommand{\msun}{{M_{\odot}}}

\newcommand{\kFn}{{k_{F_n}}}
\newcommand{\fmmo}{{\rm fm}^{-1}}
\newcommand{\fmmt}{{\rm fm}^{-3}}
\newcommand{\mev}{{\rm MeV}}
\newcommand{\mevt}{{\rm MeV/fm}^3}
\newcommand{\eos}{eos~}

\newcommand{\bag}{B^{1/4}}
\newcommand{\gcmt}{{\rm g/cm}^3}

\newcommand{\nuk}{{\nu_{\rm K}}}

\newcommand{\ergs}{{\rm erg/s}}
\newcommand{\kFchi}{k_{F_\chi}}
\def\rxj1856{\mbox{RX~J1856.5-3754}}
\def\psrotwoofive{\mbox{PSR~0205+6449}}
\newcommand{\kFp}{k_{F_p}}
\newcommand{\kFe}{k_{F_e}}
\def\sgrnineteenoo{\mbox{SGR~1900+14}}
\def\etal{{\it{et al.\/}}} 
\newcommand{\rms}{{\rm s}}

\makeindex             % used for the subject index
\begin{document}

\title*{Neutron Star Interiors and the Equation of State of Superdense Matter}
\titlerunning{Neutron Star Interiors and the eos of Superdense Matter} 

\author{Fridolin Weber\inst{1}, Rodrigo Negreiros\inst{2}, \and 
Philip Rosenfield\inst{3}}


\institute{
     San Diego State University \texttt{fweber@sciences.sdsu.edu}
\and San Diego State University \texttt{negreiro@sciences.sdsu.edu}
\and San Diego State University \texttt{philrose@sciences.sdsu.edu}
}

\maketitle

\begin{abstract}
Neutron stars contain matter in one of the densest forms found in the
Universe. This feature, together with the unprecedented progress in
observational astrophysics, makes such stars superb astrophysical
laboratories for a broad range of exciting physical studies. This
paper gives an overview of the phases of dense matter predicted to
make their appearance in the cores of neutron stars. Particular
emphasis is put on the role of strangeness. Net strangeness is carried
by hyperons, K-mesons, H-dibaryons, and strange quark matter, and may
leave its mark in the masses, radii, moment of inertia, dragging of
local inertial frames, cooling behavior, surface composition, and the
spin evolution of neutron stars.  These observables play a key role
for the exploration of the phase diagram of dense nuclear matter at
high baryon number density but low temperature, which is not
accessible to relativistic heavy ion collision experiments. 
\end{abstract}

\section{Introduction} \label{sec:1}

Neutron stars are dense, neutron-packed remnants of stars that blew apart in
supernova explosions. Many neutron stars form radio pulsars, emitting radio
waves that appear from the Earth to pulse on and off like a lighthouse beacon
as the star rotates at very high speeds. Neutron stars in x-ray binaries
accrete material from a companion star and flare to life with a burst of
x-rays.  The most rapidly rotating, currently known neutron star is pulsar PSR
J1748-2446ad, which rotates at a period of 1.39~ms (which corresponds to a
rotational frequency of 719~Hz) \cite{hessels06:a}. It is followed by PSRs
B1937+21 \cite{backer82:a} and B1957+20 \cite{fruchter88:a} whose rotational
periods are 1.58 ms (633~Hz) and 1.61~ms (621~Hz), respectively. Finally, the
recent discovery of X-ray burst oscillations from the neutron star X-ray
transient XTE J1739--285 \cite{kaaret06:a} could suggest that XTE J1739--285
contains the most rapidly rotating neutron star yet discovered.  Measurements
of radio pulsars and neutron stars in x-ray binaries comprise most of the
neutron star observations. Improved data on isolated neutron stars (e.g.
\rxj1856, \psrotwoofive) are now becoming available, and future investigations
at gravitational wave observatories focus on neutron stars as major potential
sources of gravitational waves (see \cite{villain06:a} for a recent overview).
Depending on star mass and rotational frequency, the matter in the core
regions of neutron stars may be compressed to densities that are up to an
order of magnitude greater than the density of ordinary atomic nuclei. This
extreme compression provides a high-pressure environment in which numerous
subatomic particle processes are likely to compete with each other
\cite{glen97:book,weber99:book}. The most spectacular ones stretch from the
generation of hyperons and baryon resonances ($\Sigma, \Lambda, \Xi, \Delta$),
to quark ($u, d, s$) deconfinement, to the formation of boson
\begin{figure}[tb]
\centerline{\psfig{figure=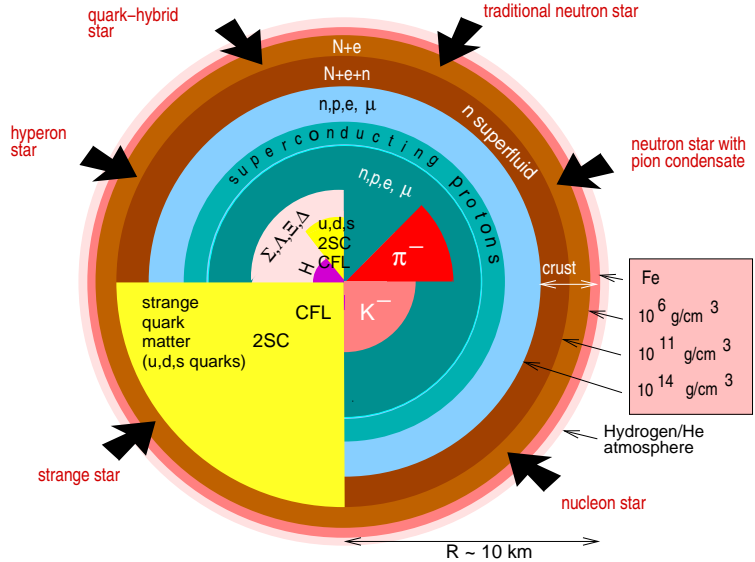,width=10.0cm}}
\caption[]{Neutron star compositions predicted by theory.}
\label{fig:crossection} 
\end{figure} 
condensates ($\pi^-$, $K^-$, H-matter)
\cite{glen97:book,weber99:book,heiselberg00:a,lattimer01:a,weber05:a,%
sedrakian06:a} (see Fig.\ \ref{fig:crossection}). In the framework of
the strange matter hypothesis \cite{bodmer71:a,witten84:a,terazawa89:a},
it has also been suggested that 3-flavor strange quark matter--made of
absolutely stable $u$, $d$, and $s$ quarks--may be more stable than
ordinary atomic nuclei. In the latter event, neutron stars should in
fact be made of such matter rather than ordinary (confined) hadronic
matter
\cite{alcock86:a,alcock88:a,madsen98:b}. Another striking implication
of the strange matter hypothesis is the possible existence of a new
class of white-dwarfs-like strange stars (strange dwarfs)
\cite{glen94:a}. The quark matter in neutron stars, strange stars, or
strange dwarfs ought to be in a color superconducting state
\cite{rajagopal01:a,alford01:a,alford98:a,rapp98+99:a}. This
fascinating possibility has renewed tremendous interest in the physics
of neutron stars and the physics and astrophysics of (strange) quark
matter \cite{weber05:a,rajagopal01:a,alford01:a}.  This paper
discusses the possible phases of ultra-dense nuclear matter expected
to exist deep inside neutron stars and reviews selected models derived
for the equation of state (eos) of such matter (see also
\cite{glen97:book,weber99:book,heiselberg00:a,lattimer01:a,blaschke01:trento,%
weber05:a,sedrakian06:a,page06:review,blaschke06:a}).

\section{Neutron Star Masses}\label{sec:masses}

\begin{figure}[tb]
\centerline{\psfig{figure=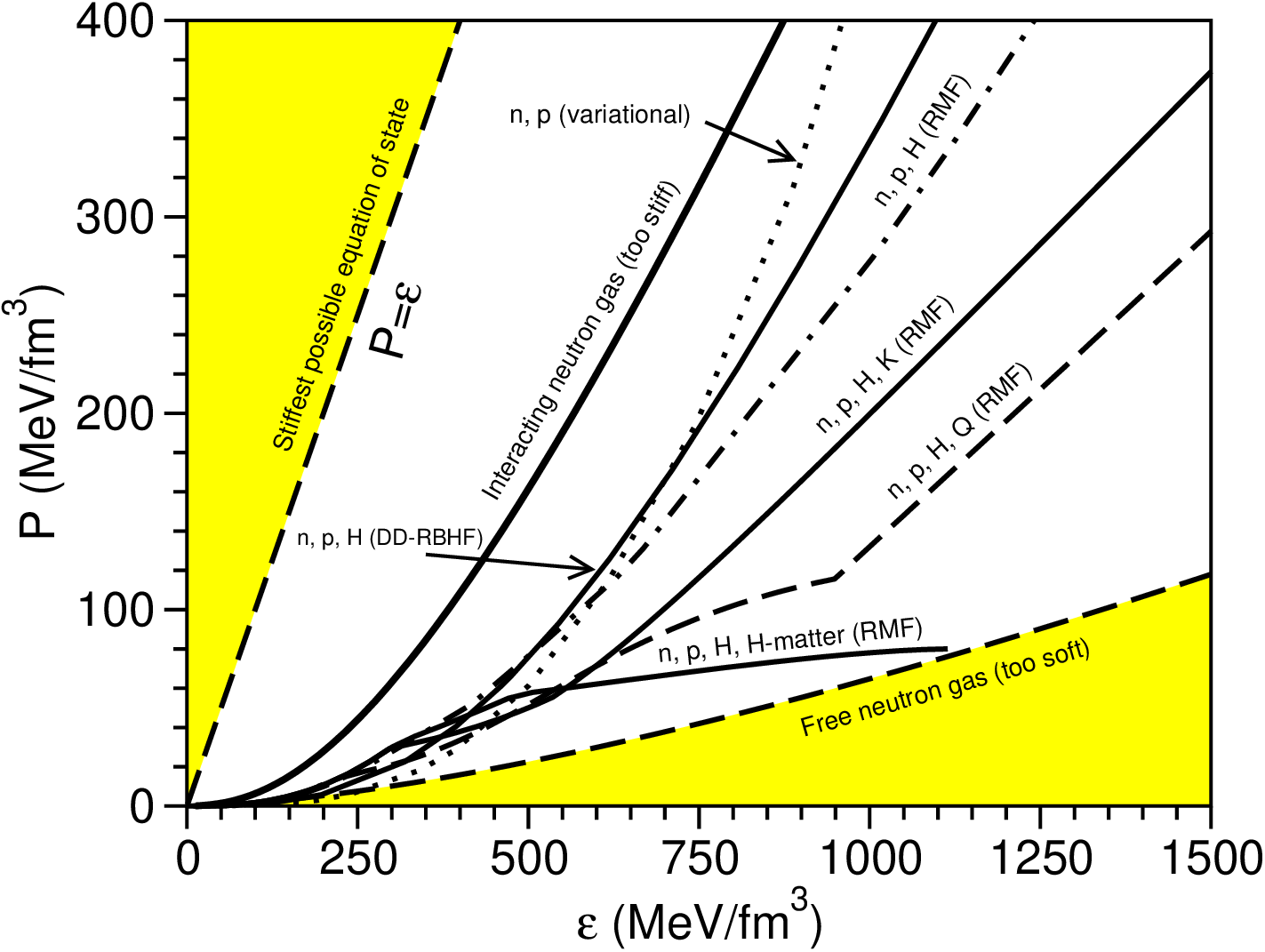,width=10.0cm}}
\caption[]{Models for the equation of state (pressure versus energy 
density) of neutron star matter \cite{weber05:a}. The notation is as
follows: RMF=relativistic mean-field model; DD-RBHF=density dependent
relativistic Brueckner-Hartree-Fock model; n=neutrons; p=protons;
H=hyperons, K=$K^-[u,\bar s]$ meson condensate; Q=$u,d,s$ quarks;
H-matter=H-dibaryon condensate.}
\label{fig:nucl_eoss} 
\end{figure}

\begin{figure}[tb]
\centerline{\psfig{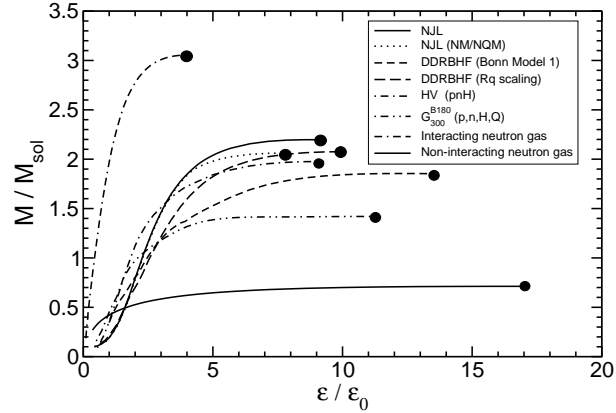}}
\caption{Neutron star mass versus central density (in units of
nuclear matter density, $\epsilon_0 = 140~\mevt$).}
\label{fig:ec} 
\end{figure}

In 1939, Tolman, Oppenheimer and Volkoff performed the first
neutron star calculations, assuming that such objects are entirely
made of a gas of non-interacting relativistic neutrons
\cite{oppenheimer39,tolman39:a}. The \eos of such a gas is extremely
soft, i.e.\ very little additional pressure is gained with increasing
density, as can be seen from Fig.\
\ref{fig:nucl_eoss}, and predicts a
maximum neutron star mass of just $0.7~\msun$ (Fig.\ \ref{fig:MvsR})
at an unrealistically high density of 17 times the density of nuclear
matter (Fig.\ \ref{fig:ec}).  It is interesting to note that the
inclusion of interactions among the neutrons increases the star's maximum
mass from $0.7~\msun$ to around $3~
\msun$ (Figs.\ \ref{fig:MvsR} and \ref{fig:ec}). However, the radii of 
the latter are so big that mass shedding from the star's equator
occurs at rotational frequencies that are considerably smaller than
those observed for PSR~J1748-2446ad, 716~Hz (1.39~ms)
\cite{hessels06:a}, or B1937+21, 630~Hz (1.58~ms)
\cite{backer82:a}. An interacting neutron gas thus fails to
accommodate the observed rapidly rotating neutron stars.  The other
extreme, a non-interacting relativistic neutron gas, fails too since
it does not accommodate the Hulse-Taylor pulsar ($M=1.44\, \msun$)
\cite{taylor89:a}, and also conflicts with the average neutron star
mass of $1.350 \pm 0.004\, \msun$ derived by Thorsett and Chakrabarty
\cite{thorsett99:a} from observations of radio pulsar systems.  More
than that, recent observations indicate that neutron star masses may
be as high as around $2~ \msun$. Examples of such very heavy neutron
stars are $M_{\rm J0751+1807} = 2.1 \pm 0.2 ~
\msun$ \cite{nice05:b}, $M_{\rm 4U\,1636+536} = 2.0 \pm 0.1~
\msun$ \cite{barret06:a}, $M_{\rm Vela\, X-1} = 1.86\pm 0.16\,
\msun$ \cite{barziv01:a}, $M_{\rm Cyg\, X-2} = 1.78\pm 0.23 \,
\msun$ \cite{casares98:a,orosz99:a}.  Large masses have also been
reported for the high-mass x-ray binary 4U\,1700--37 and the compact
object in the low-mass x-ray binary 2S0921--630, $M_{\rm 4U\,1700-37}
= 2.44 \pm 0.27~ \msun$ \cite{clark02:a} and $M_{\rm 2S0921-630} = 2.0
- 4.3 \msun$ \cite{shahbaz04:a}. respectively. The latter two objects
may be either massive neutron stars or low-mass black holes with
masses slightly higher than the maximum possible neutron star mass of
$~\sim 3 \msun$. This value follows from a general, theoretical
estimate of the maximal possible mass of a stable neutron star as
performed by Rhoades and Ruffini \cite{rhoades74:a} on the basis that
(1) Einstein's theory of general relativity is the correct theory of
gravity, (2) the \eos satisfies both the microscopic stability
condition $\partial P/\partial\epsilon \geq 0$ and the causality
condition $\partial P / \partial\epsilon \leq c^2$, and (3) that the
\eos below some matching density is known. From these assumptions, it
follows that the maximum mass of the equilibrium configuration of a
neutron star cannot be larger than $3.2\,\msun$.  This value increases
to about $5\,\msun$ if one abandons the causality constraint $\partial
P/\partial\epsilon\leq c^2$ \cite{sabbadini77:a,hartle78:a}, since it
allows the \eos to behave stiffer at asymptotically high nuclear
densities.  If either one of the two objects 4U\,1700--37 or
2S0921--630 were a black hole, it would confirm the prediction of the
existence of low-mass black holes \cite{brown94:a}. Conversely, if
these objects were massive neutron stars, their high masses would
severely constrain the \eos of dense nuclear matter. 

\section{Composition of Cold and Dense Neutron Star Matter}

\begin{figure}[tb]
\centerline{\psfig{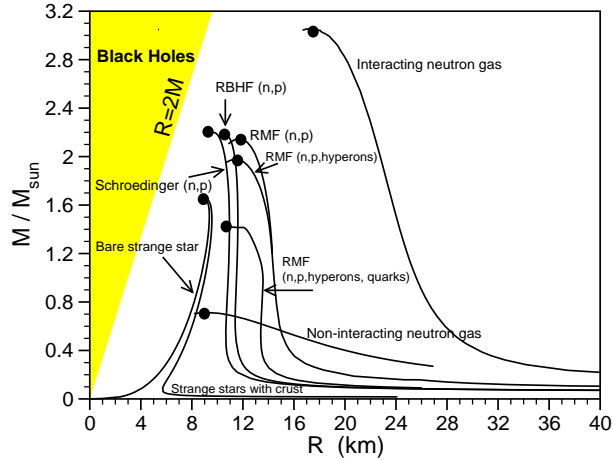}}
\caption[]{Mass-radius relationship of neutron stars and strange
stars \cite{weber05:a}. The strange stars may be enveloped in a crust
of ordinary nuclear material whose density is below neutron drip
density \cite{alcock86:a,glen92:crust,stejner05:a}.}
\label{fig:MvsR} 
\end{figure}

A vast number of models for the equation of state of neutron star
matter has been derived in the literature over the years. These models
can roughly be classified as follows:
\begin{itemize}
\item Thomas-Fermi based models \cite{myers95:a,strobel97:a}
\item Schroedinger-based models (e.g.\
variational approach, Monte Carlo techniques, hole line expansion
(Brueckner theory), coupled cluster method, Green function method)
\cite{heiselberg00:a,pandharipande79:a,wiringa88:a,akmal98:a,%
baldo99:BBG,baldo01:springer,burgio02:a}
\item Relativistic field-theoretical treatments (relativistic mean field
(RMF), Hartree-Fock (RHF), standard Brueckner-Hartree-Fock (RBHF),
density dependent RBHF (DD-RBHF)
\cite{lenske95:a,fuchs95:a,typel99:a,hofmann01:a,niksic02:a,ban04:a}
\item Nambu-Jona-Lasinio (NJL) models 
\cite{buballa05:a,blaschke05:a,rischke05:a,abuki06:a,lawley06:a,lawley06:b}
\item Chiral SU(3) quark mean field model \cite{wang05:a}.
\end{itemize}
A collection of equations of state computed for several of these
models is shown in Fig.\ \ref{fig:nucl_eoss}. Mass--radius
relationships of neutron stars based on these equations of
state are shown in Fig.\ \ref{fig:MvsR}.
Any acceptable nuclear many-body calculation must correctly reproduce
the bulk properties of nuclear matter at saturation density, $n_0 =
0.16~\fmmt$.  These are the binding energy, $E/A=-16.0$~MeV, effective
nucleon mass, $m^*_{\rm N} = 0.79 \,m_{\rm N}$, incompressibility,
$K\simeq 240$~MeV, and the symmetry energy, $a_{\rm s}= 32.5$~MeV.

\goodbreak
\subsection{Hyperons and baryon resonances}

\begin{figure}[tb]
\begin{center}
\parbox[t]{5.7cm} {\psfig{file=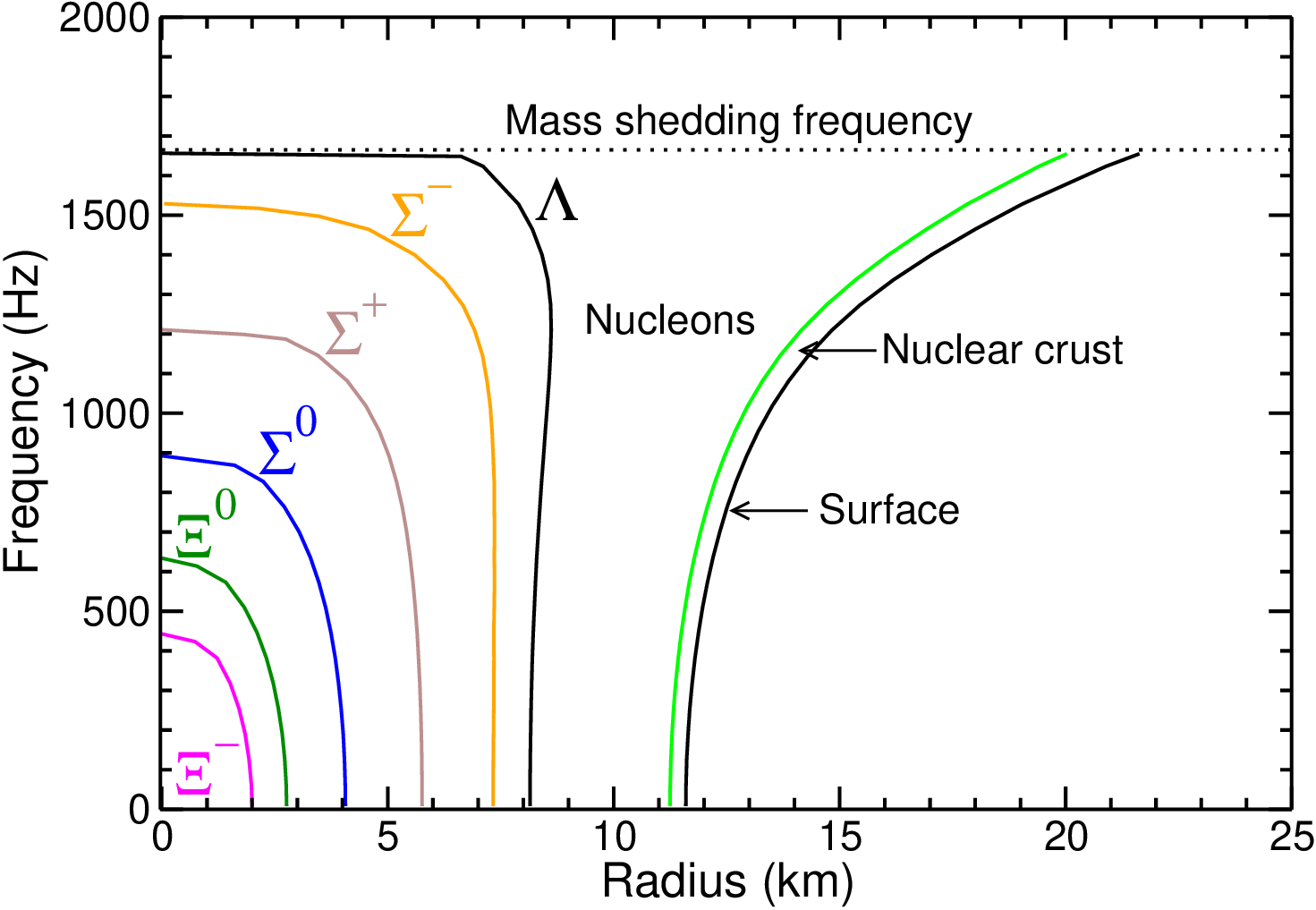,width=5.7cm}
{\caption[]{Hyperon composition of a rotating neutron star in
equatorial direction.}
\label{fig:bonn_eq}}}
\ \hskip 0.1cm \
\parbox[t]{5.7cm}
{\psfig{file=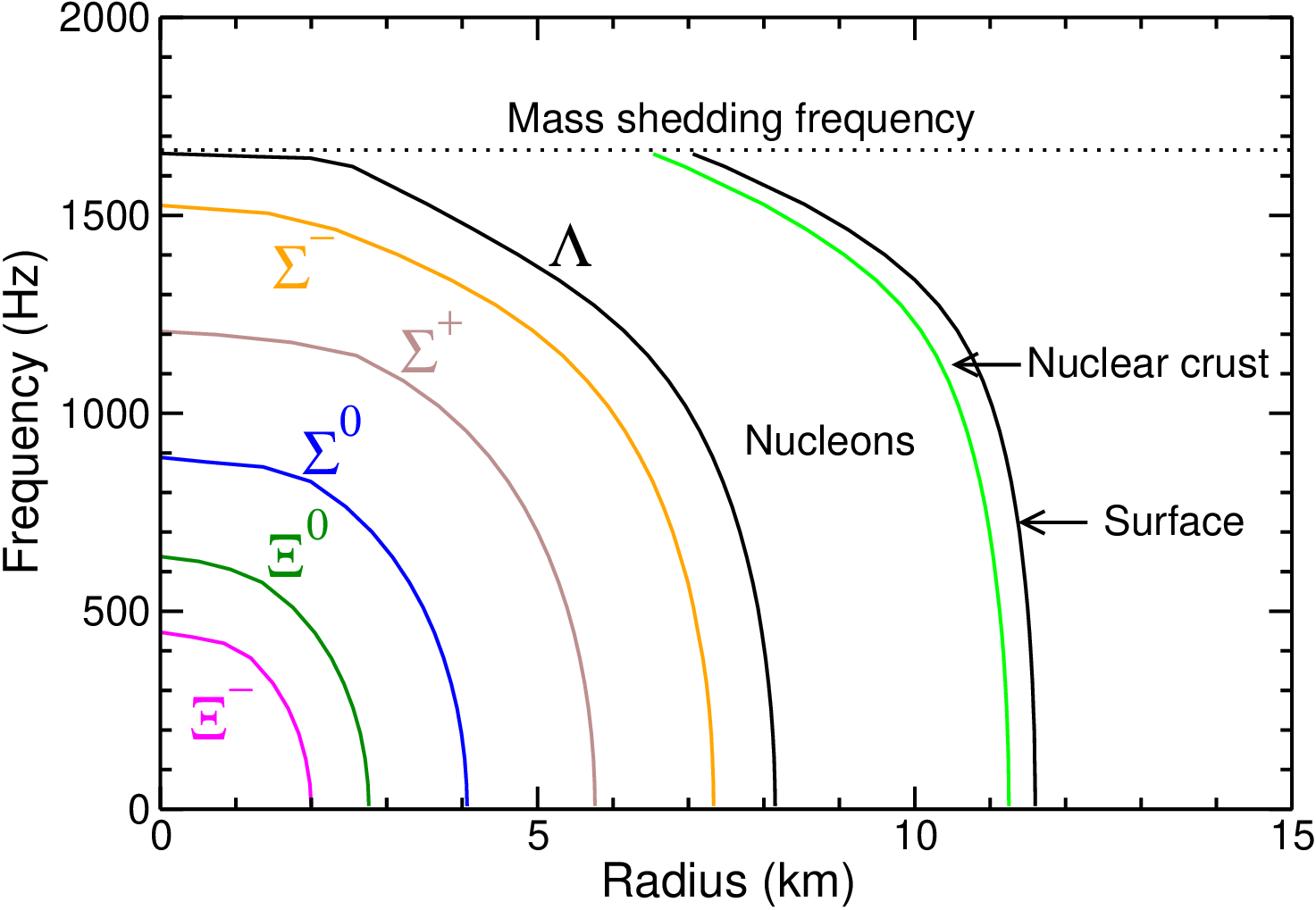,width=5.7cm} {\caption[]{
Same as Fig.\ \ref{fig:bonn_eq}, but in polar direction.}
\label{fig:bonn_po}}}
\end{center}
\end{figure}

At the densities in the interior of neutron stars, the neutron
chemical potential, $\mu^n$, is likely to exceed the masses, modified
by interactions, of $\Sigma,~ \Lambda$ and possibly $\Xi$ hyperons
\cite{glen85:b}. Hence, in addition to nucleons, neutron star matter
may be expected to contain significant populations of strangeness
carrying hyperons.  The thresholds of the lightest baryon resonances
($\Delta^-, \Delta^0, \Delta^+, \Delta^{++}$) are not reached in
relativistic mean-field (Hartree) calculations. This is different for
many-body calculations performed at the relativistic
Brueckner-Hartree-Fock level, where $\Delta$'s appear very abundantly
\cite{huber98:a}. In any event, pure neutron matter constitutes an
excited state relative to hyperonic matter which, therefore, would
quickly transform via weak reactions like
\begin{equation}
n \rightarrow p + e^- + {\bar{\nu}}_e 
\label{eq:cnp}
\end{equation} to the lower energy state. The chemical 
potentials associated with reaction (\ref{eq:cnp}) in equilibrium obey the
relation
\begin{equation}
\mu^n = \mu^p + \mu^{e^-} \, ,
\label{eq:mun}
\end{equation} 
where $\mu^{\bar\nu_e}=0$ since the mean free path of (anti) neutrinos
is much smaller than the radius of neutron stars. Hence (anti)
neutrinos do not accumulate inside neutron stars. This is different
for hot proto-neutron stars \cite{prakash97:a}.
Equation~(\ref{eq:mun}) is a special case of the general relation
\begin{equation}
\mu^\chi = B^\chi \mu^n - q^\chi \mu^{e^-} \, , 
\label{eq:mui}
\end{equation} which holds in any system characterized by two conserved
charges. These are in the case of neutron star matter electric charge,
$q^\chi$, and baryon number charge, $B^\chi$. Application of Eq.\
(\ref{eq:mui}) to the $\Lambda$ hyperon ($B^\Lambda=1$, $q^\Lambda=0$), for
instance, leads to
\begin{equation}
\mu^\Lambda = \mu^n \, .
\label{eq:L}
\end{equation} Ignoring particle interactions, the chemical potential of a
relativistic particle of type $\chi$ is given by 
\begin{equation}
\mu^\chi = \omega(\kFchi) \equiv \sqrt{m_\chi^2 + \kFchi^2} \, ,
\label{eq:mukF}
\end{equation} where $\omega(\kFchi)$ is the single-particle 
energy of the particle and $\kFchi$ its Fermi momentum.  Substituting
(\ref{eq:mukF}) into (\ref{eq:L}) leads to
\begin{equation}
  \kFn \geq \sqrt{m_\Lambda^2 - m_n^2} \simeq 3~\fmmo \Rightarrow
    n \equiv { {\kFn^3}\over{3 \pi^2} } \simeq 6 n_0 \, ,
\label{eq:kFn}
\end{equation} 
where $m_\Lambda=1116$~MeV and $m_n=939$~MeV was used. That is, if
interactions among the particles are ignored, neutrons are replaced with
$\Lambda$'s in neutron star matter at densities of around six times the
density of nuclear matter. This value is reduced to about $2 \, n_0$ by the
inclusion of particle interactions \cite{glen85:b}.  Densities that small are
easily reached in the cores of neutron stars. Hence, in addition to nucleons
and electrons, neutron stars may be expected to contain considerable
populations of strangeness-carrying $\Lambda$ hyperons, possibly accompanied
by smaller populations of the charged states of the $\Sigma$ and $\Xi$
hyperons \cite{glen85:b}. Depending on the star's mass, the total hyperon
population can be very large \cite{glen85:b}, which is illustrated graphically
in Figs.\ \ref{fig:bonn_eq} and \ref{fig:bonn_po} for rotating neutron stars
whose equation of state is computed in the framework of the relativistic
DD-RBHF formalism \cite{hofmann01:a}.  Aside from chemical equilibrium, the
condition of electric charge neutrality of neutron star matter,
\begin{equation}
  \sum_{\chi=p, \Sigma^\pm, \Xi^-, \Delta^{++} , ...;
 e^-, \mu^-} q^\chi ~ {\kFchi^3} ~+ ~ 3 \, \pi^2 \, n^M \,
 \Theta(\mu^M - m_M) \equiv 0 \, ,
\label{eq:charge2}
\end{equation}
where $M$ stands for $\pi^-$ or $K^-$ mesons, plays a key role for the
particle composition of neutron star matter too. The last term in
(\ref{eq:charge2}) accounts for the possible existence of either a
$\pi^-$ or a $K^-$ meson condensate in neutron star matter, which will
be discussed in more detail in Sect.\ \ref{ssec:mcondens}. Before,
however, we illustrate the importance of Eqs.\ (\ref{eq:mun}) and
(\ref{eq:charge2}) for the proton-neutron fraction of neutron star
matter.  The beta decay and electron capture processes among nucleons,
$n \rightarrow p + e^- + \bar{\nu}_e$ and $p + e^-
\rightarrow n + \nu_e$ respectively, also known as nucleon direct Urca
processes, are only possible in neutron star matter if the proton
fraction exceeds a certain critical value
\cite{lattimer91:a}. Otherwise energy and momentum can not be
conserved simultaneously for these reactions so that they are
forbidden. For a neutron star made up of only nucleons and electrons,
it is rather straightforward to show that the critical proton fraction
is around $11\%$. This follows from ${\mathbf k}_{F_n} = {\mathbf
k}_{F_p} + {\mathbf k}_{F_e}$ combined with the condition of electric
charge neutrality of neutron star matter. The triangle inequality then
requires for the magnitudes of the particle Fermi momenta $k_{F_n}
\leq k_{F_p} + k_{F_e}$, and charge neutrality dictates that $\kFp =
\kFe$.  Substituting $\kFp = \kFe$ into the triangle inequality leads
to $\kFn \leq 2 \kFp$ so that for the particle number densities of
neutrons and protons $n_n \leq 8 n_p$.  Expressed as a fraction of the
system's total baryon number density, $n\equiv n_p + n_n$, one thus
arrives at $n_p / n > 1/9 \simeq 0.11$, which is the figure quoted
just above.  Medium effects and interactions among the particles
modify this value only slightly but the presence of muons raises it to
about $0.15$.  Hyperons, which may exist in neutron star matter rather
abundantly, produce neutrinos via direct Urca processes like $\Sigma^-
\rightarrow \Lambda + e^- + \bar{\nu}_e$ and $\Lambda + e^-
\rightarrow \Sigma^- + \nu_e$ \cite{prakash92:a}.  The direct Urca
processes are of key importance for neutron star cooling (see D.\
Page's contribution elsewhere in this volume).  In most cases, the
nucleon direct Urca process is more efficient than the ones involving
hyperons \cite{haensel94:a,schaab95:a}.

\goodbreak
\subsection{Meson condensation}\label{ssec:mcondens}

The condensation of negatively charged mesons in neutron star matter
is favored because such mesons would replace electrons with very high
Fermi momenta. Early estimates predicted the onset of a negatively
charged pion condensate at around $2 n_0$ (see, for instance, Ref.\
\cite{baym78:a}). However, these estimates are very sensitive to
the strength of the effective nucleon particle-hole repulsion in the
isospin $T=1$, spin $S=1$ channel, described by the Landau
Fermi-liquid parameter $g'$, which tends to suppress the condensation
mechanism. Measurements in nuclei tend to indicate
that the repulsion is too strong to permit condensation in nuclear
matter \cite{barshay73:a,brown88:a}. In the mid 1980s, it was
discovered that the in-medium properties of $K^- [u \bar s]$ mesons
may be such that this meson rather than the $\pi^-$ meson may condense
in neutron star matter \cite{kaplan86:a,brown87:a,lee95:a}.

The condensation is initiated by the schematic reaction $e^-
\rightarrow K^- + \nu_e$.  If this reaction becomes possible in
neutron star matter, it is energetically advantageous to replace the
fermionic electrons with the bosonic $K^-$ mesons. Whether or not this
happens depends on the behavior of the $K^-$ mass, $m^*_{K^-}$, in
neutron star matter.  Experiments which shed light on the properties
of the $K^-$ in nuclear matter have been performed with the Kaon
Spectrometer (KaoS) and the FOPI detector at the heavy-ion synchrotron
SIS at GSI
\cite{barth97:a,senger01:a,sturm01:a,devismes02:a,fuchs06:a}.  An
analysis of the early $K^-$ kinetic energy spectra extracted from
Ni+Ni collisions \cite{barth97:a} showed that the attraction from
nuclear matter would bring the $K^-$ mass down to $m^*_{K^-}\simeq
200~\mev$ at densities $\sim 3\, n_0$. For neutron-rich matter, the
relation $m^*_{K^-} / m_{K^-} \simeq 1 - 0.2 n / n_0$ was established
\cite{li97:a,li97:b,brown97:a}, with $m_K = 495$~MeV the $K^-$ vacuum
mass.  Values of around $m^*_{K^-}\simeq 200~\mev$ may be reached by
the electron chemical potential, $\mu^e$, in neutron star matter
\cite{weber99:book,glen85:b} so that the threshold condition for the
onset of $K^-$ condensation, $\mu^e = m^*_K$ might be fulfilled for
sufficiently dense neutron stars, provided other negatively charged
particles ($\Sigma^-$, $\Delta^-$, $d$ and $s$ quarks) are not
populated first and prevent the electron chemical potential from
increasing with density.

We also note that $K^-$ condensation allows the conversion reaction $n
\rightarrow p + K^-$. By this conversion the nucleons in the cores of
neutron stars can become half neutrons and half protons, which lowers
the energy per baryon of the matter \cite{brown96:a}. The relative
isospin symmetric composition achieved in this way resembles the one
of atomic nuclei, which are made up of roughly equal numbers of
neutrons and protons.  Neutron stars are therefore referred to, in
this picture, as nucleon stars. The maximum mass of such stars has
been calculated to be around $1.5\,
\msun$ \cite{thorsson94:a}. Consequently, the collapsing core of a
supernova, e.g.\ 1987A, if heavier than this value, should go into a
black hole rather than forming a neutron star, as pointed out by Brown
et al.\ \cite{brown94:a,li97:a,li97:b}. This would imply the existence of
a large number of low-mass black holes in our galaxy \cite{brown94:a}.
Thielemann and Hashimoto \cite{thielemann90:a} deduced from the total
amount of ejected $^{56}{\rm Ni}$ in supernova 1987A a neutron star
mass range of $1.43 - 1.52~ \msun$. If the maximum neutron star mass
should indeed be in this mass range ($\sim 1.5~\msun$),  
the existence of heavy neutron stars with masses
around $2~\msun$ (Sect.\ \ref{sec:masses}) would be ruled out. 
Lastly, we mention that meson condensates lead to neutrino
luminosities which are considerably enhanced over those of normal
neutron star matter. This would speed up neutron star cooling
considerably \cite{thorsson94:a,schaab95:a}.

\subsection{H-matter and exotic baryons}

A novel particle that could be of relevance for the composition of
neutron star matter is the H-dibaryon (H=$([ud][ds][su])$), a doubly
strange six-quark composite with spin and isospin zero, and baryon
number two \cite{jaffe77:a}. Since its first prediction in 1977, the
H-dibaryon has been the subject of many theoretical and experimental
studies as a possible candidate for a strongly bound exotic state. In
neutron star matter, which may contain a significant fraction of
$\Lambda$ hyperons, the $\Lambda$'s could combine to form H-dibaryons,
which could give way to the formation of H-dibaryon matter at
densities somewhere above $\sim 4\,
n_0$ \cite{tamagaki91:a,sakai97:a,glen98:a}.  If formed in neutron
stars, however, H-matter appears to unstable against compression which
could trigger the conversion of neutron stars into hypothetical
strange stars \cite{sakai97:a,faessler97:a,faessler97:b}.

Another particle, referred to as exotic baryon, of potential relevance
for neutron stars, could be the pentaquark, $\Theta^+ ([ud]^2 \bar
s)$, with a predicted mass of 1540~MeV. The pentaquark, which carries
baryon number one, is a hypothetical subatomic particle consisting of
a group of four quarks and one anti-quark (compared to three quarks in
normal baryons and two in mesons), bound by the strong color-spin
correlation force (attraction between quarks in the color $\bar {\bf
3}_c$ channel) that drives color
superconductivity \cite{jaffe03:a,jaffe05:a}. The pentaquark decays
according to $\Theta^+(1540) \rightarrow K^+ [\bar s u] + n[udd]$ and
thus has the same quantum numbers as the $K^+ n$. The associated
reaction in chemically equilibrated matter would imply $\mu^{\Theta^+}
= \mu^{K^+} + \mu^n$.

\subsection{Quark deconfinement}\label{ssec:deconf}

It has been suggested already many decades ago
\cite{ivanenko65:a,itoh70:a,fritzsch73:a,baym76:a,keister76:a,%
  chap77:a,fech78:a,chap77:b} that the nucleons may melt under the enormous
pressure that exists in the cores of neutron stars, creating a new state of
matter know as quark matter. From simple geometrical considerations it follows
that for a characteristic nucleon radius of $r_N\sim 1$~fm, nucleons may begin
to touch each other in nuclear matter at densities around $(4\pi r^3_N/3)^{-1}
\simeq 0.24~\fmmt = 1.5\, n_0$, which is less than twice the density of
nuclear matter. This figure increases to $\sim 11 \, n_0$ for a nucleon radius
of $r_N = 0.5$~fm. One may thus speculate that the hadrons of neutron star
matter begin to dissolve at densities somewhere between around $2-10\, n_0$,
giving way to unconfined quarks.  Depending on rotational frequency and
neutron star mass, densities greater than two to three times $n_0$ are easily
reached in the cores of neutron stars so that the neutrons and protons in the
cores of neutron stars may indeed be broken up into their quark constituents
\cite{glen97:book,weber99:book,weber05:a,glen91:pt}. More than that, since the
mass of the strange quark is only $m_s \sim 150$~MeV, high-energetic up and
down quarks will readily transform to strange quarks at about the same density
at which up and down quark deconfinement sets in. Thus, if quark matter exists
in the cores of neutron stars, it should be made of the three lightest quark
flavors.  A possible astrophysical signal of quark deconfinement in the cores
of neutron stars was suggested in \cite{glen97:a}.  The remaining three quark
flavors (charm, top, bottom) are way to massive to be created in neutron
stars. For instance, the creation of charm quark requires a density greater
than $10^{17}\, \gcmt$, which is around 100 times greater than the density
reached in neutron stars.  A stability analysis of stars with a charm quark
population reveals that such objects are unstable against radial oscillations
and, thus, can not exist stably in the universe \cite{weber99:book,weber05:a}.
The same is true for ultra-compact stars with unconfined populations of top
and bottom quarks, since the pulsation eigen-equations are of Sturm-Liouville
type.

The phase transition from confined hadronic (H) matter to deconfined
quark (Q) matter is characterized by the conservation of baryon charge
and electric charge. The Gibbs condition for phase equilibrium then is
that the two associated chemical potentials, $\mu^n$ and $\mu^e$, and
the pressure in the two phases be equal \cite{glen97:book,glen91:pt},
\begin{eqnarray}
  P_{\rm H}(\mu^n,\mu^e, \{ \chi \}, T) = P_{\rm Q}(\mu^n,\mu^e,T) \,
  ,
\label{eq:gibbs1}
\end{eqnarray} The quantity $P_{\rm H}$ denotes the pressure of hadronic
matter computed for a given hadronic Lagrangian ${\cal L}_{\rm
M}(\{\chi\})$, where $\{\chi\}$ denotes the field variables and Fermi
momenta that characterize a solution to the field equations of
confined hadronic matter,
\begin{eqnarray}
  ( i \gamma^\mu\partial_\mu - m_\chi ) \psi_\chi(x) &=&
  \sum_{M=\sigma,\omega,\pi, ...} \Gamma_{M \chi} M(x) \, \psi_\chi(x) \, ,
   \\ 
 ( \partial^\mu\partial_\mu + m^2_\sigma) \sigma(x) &=&
  \sum_{\chi = p, n, \Sigma, ...} \Gamma_{\sigma \chi}\, \bar\psi_\chi(x)
  \psi_\chi(x) \, , 
\end{eqnarray}
 plus additional equations for the other meson fields ($M= \omega,
\pi, \rho , ...$).  The pressure of quark matter, $P_{\rm Q}$, is
obtainable from the bag model \cite{chodos74:a,chodos74:b}. The quark
chemical potentials $\mu^u, ~\mu^d, ~\mu^s$ are related to the baryon
and charge chemical potentials as
\begin{eqnarray}
  \mu^u = {{1}\over{3}} \, \mu^n - {{2}\over{3}} \, \mu^e\,
  ,\qquad \mu^d = \mu^s = {{1}\over{3}} \, \mu^n + {{1}\over{3}} \,
  \mu^e \, .
\label{eq:cp.ChT}
\end{eqnarray} Equation~(\ref{eq:gibbs1}) is to be supplemented with
the two global relations for conservation of baryon charge and
electric charge within an unknown volume $V$ containing $A$
baryons. The first one is given by
\begin{equation}
  n \equiv {A\over V} = (1-\eta) \, n_{\rm H}(\mu^n,\mu^e,T) +
  \eta \, n_{\rm Q}(\mu^n,\mu^e,T) \, ,
\label{eq:bcharge}
\end{equation} where $\eta \equiv V_{\rm Q}/V$ denotes the volume
proportion of quark matter, $V_{\rm Q}$, in the unknown volume $V$, and
$n_{\rm H}$ and $n_{\rm Q} $ are the baryon number densities of hadronic
matter and quark matter.  Global neutrality of electric charge within the
volume $V$ can be written as
\begin{equation}
  0 = {Q\over V} = (1-\eta) \, q_{\rm H}(\mu^n,\mu^e,T) + \eta \,
  q_{\rm Q}(\mu^n,\mu^e,T)  +  q_{\rm L} \, ,
\label{eq:echarge}
\end{equation} with $q_i$ the electric charge densities of
hadrons, quarks, and leptons.  For a given temperature, $T$, Eqs.\
(\ref{eq:gibbs1}) to (\ref{eq:echarge}) serve to determine the two
independent chemical potentials and the volume $V$ for a specified
\begin{figure}[tb]
\begin{center}
\includegraphics*[width=0.9\textwidth,angle=0,clip]{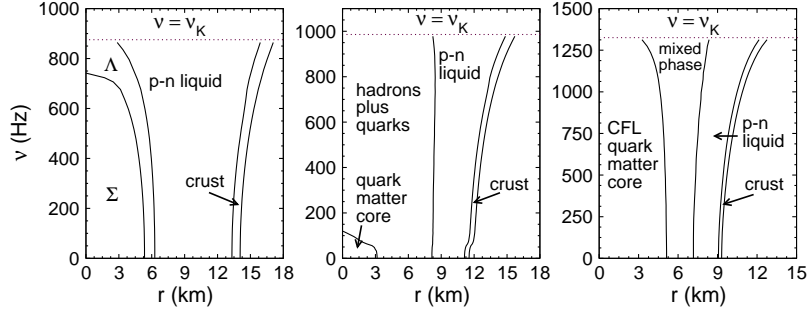}
\caption{Dependence of neutron star composition on spin frequency, $\nu$, 
for three sample compositions (left: hyperon composition, middle:
quark-hybrid composition, right: quark-hybrid composition with quark
matter in the color-flavor locked (CFL) phase \cite{weber06:a}). The
non-rotating stellar mass in each case is  $1.4\, \msun$. $\nuk$
denotes the Kepler (mass-shedding) frequency, which sets an absolute
limit on stable rotation.}
\label{fig:profiles}
\end{center}
\end{figure}
volume fraction $\eta$ of the quark phase in equilibrium with the
hadronic phase.  After completion $V_{\rm Q}$ is obtained as $V_{\rm
Q}=\eta V$. Because of Eqs.\ (\ref{eq:gibbs1}) through
(\ref{eq:echarge}) the chemical potentials depend on the proportion
$\eta$ of the phases in equilibrium, and hence so also all properties
that depend on them, i.e.\ the energy densities, baryon and charge
densities of each phase, and the common pressure. For the mixed phase,
the volume proportion of quark matter varies from $0 \leq \eta \leq
1$ and the energy density is the linear combination of the two
phases \cite{glen97:book,glen91:pt},
\begin{equation}
  \epsilon = (1-\eta) \, \epsilon_{\rm H}(\mu^n,\mu^e, \{\chi\}, T) + \eta \,
  \epsilon_{\rm Q}(\mu^n,\mu^e,T) \, .
\label{eq:eps.chi}
\end{equation} Hypothetical neutron star compositions computed 
along the lines described above are shown in Fig.\ \ref{fig:profiles}.
Possible astrophysical signals associated with quark deconfinement,
the most striking of which being ``backbending'' of isolated pulsars,
are discussed in
\cite{glen97:book,weber99:book,weber05:a,weber99:topr,glen00:b}.

\goodbreak
\subsection{Color-superconductivity}\label{sec:color}

There has been much recent progress in our understanding of quark
matter, culminating in the discovery that if quark matter exists it
ought to be in a color superconducting state
\cite{rajagopal01:a,alford01:a,alford98:a,rapp98+99:a}. This is made
possible by the strong interaction among the quarks which is very
attractive in some channels. Pairs of quarks are thus expected to form
Cooper pairs very readily. Since pairs of quarks cannot be
color-neutral, the resulting condensate will break the local color
symmetry and form what is called a color superconductor.  The phase
diagram of such matter is expected to be very complex
\cite{rajagopal01:a,alford01:a}. The complexity is caused by the fact that
quarks come in three different colors, different flavors, and
different masses.  Moreover, bulk matter is neutral with respect to
both electric and color charge, and is in chemical equilibrium under
the weak interaction processes that turn one quark flavor into
another. To illustrate the condensation pattern briefly, we note the
following pairing ansatz for the quark condensate \cite{alford03:a},
\begin{eqnarray}
\langle \psi^\alpha_{f_a} C\gamma_5 \psi^\beta_{f_b} \rangle \sim
\Delta_1 \epsilon^{\alpha\beta 1}\epsilon_{{f_a}{f_b}1} + \Delta_2
\epsilon^{\alpha\beta 2}\epsilon_{{f_a}{f_b}2} + \Delta_3
\epsilon^{\alpha\beta 3}\epsilon_{{f_a}{f_b}3} \, ,
\label{eq:pairing_ansatz}
\end{eqnarray}
where $\psi^\alpha_{f_a}$ is a quark of color $\alpha=(r,g,b)$ and
flavor ${f_a}=(u,d,s)$. The condensate is a Lorentz scalar,
antisymmetric in Dirac indices, antisymmetric in color, and thus
antisymmetric in flavor. The gap parameters $\Delta_1$, $\Delta_2$ and
$\Delta_3$ describe $d$-$s$, $u$-$s$ and $u$-$d$ quark Cooper pairs,
respectively. The following pairing schemes have emerged. At
asymptotic densities ($m_s \rightarrow 0$ or $\mu \rightarrow \infty$)
the ground state of QCD with a vanishing strange quark mass is the
color-flavor locked (CFL) phase (color-flavor locked quark pairing),
in which all three quark flavors participate symmetrically.  The
gaps associated with this phase  are
\begin{equation}
\Delta_3 \simeq \Delta_2 = \Delta_1 = \Delta \, ,
\label{eq:delta_CFL}
\end{equation}
and the quark condensates of the CFL phase are approximately of the form
\begin{equation}
 \langle \psi^{\alpha}_{f_a} C\gamma_5 \psi^{\beta}_{f_b} \rangle
\sim \Delta \, \epsilon^{\alpha \beta X} \epsilon_{{f_a} {f_b} X}
\, ,
\label{eq:CFL1}
\end{equation}
with color and flavor indices all running from 1 to 3. Since
$\epsilon^{\alpha\beta X} \epsilon_{{f_a} {f_b} X} =
\delta^\alpha_{f_a}\delta^\beta_{f_b} -
\delta^\alpha_{f_b}\delta^\beta_{f_a}$ one sees that the condensate
(\ref{eq:CFL1}) involves Kronecker delta functions that link color and
flavor indices. Hence the notion color-flavor locking. The CFL phase
has been shown to be electrically neutral without any need for
electrons for a significant range of chemical potentials and strange
quark masses \cite{rajagopal01:b}. If the strange quark mass is heavy
enough to be ignored, then up and down quarks may pair in the
two-flavor superconducting (2SC) phase.  Other possible condensation
patterns are CFL-$K^0$ \cite{bedaque01:a}, CFL-$K^+$ and
CFL-$\pi^{0,-}$ \cite{kaplan02:a}, gCFL (gapless CFL phase)
 \cite{alford03:a}, 1SC (single-flavor-pairing)
 \cite{alford03:a,buballa02:a,schmitt04:a}, CSL (color-spin locked
phase)  \cite{schaefer00:a}, and the LOFF (crystalline pairing)
 \cite{alford00:a,bowers02:a,casalbuoni04:a} phase, depending on $m_s$,
$\mu$, and electric charge density.  Calculations performed for
massless up and down quarks and a very heavy strange quark mass ($m_s
\rightarrow \infty$) agree that the quarks prefer to pair in the
two-flavor superconducting (2SC) phase where
\begin{equation}
\Delta_3 > 0\, , \quad {\rm and} \quad  \Delta_2 = \Delta_1 = 0 \, .
\label{eq:delta_2SC}
\end{equation}
In this case the pairing ansatz (\ref{eq:pairing_ansatz}) reduces to
\begin{equation}
 \langle \psi^{\alpha}_{f_a} C \gamma_5 \psi^{\beta}_{f_b} \rangle
\propto \Delta \, \epsilon_{ab} \epsilon^{\alpha \beta 3} \, .
\label{eq:2SC}
\end{equation}
Here the resulting condensate picks a color direction (3 or blue in
the example (\ref{eq:2SC}) above), and creates a gap $\Delta$ at the
Fermi surfaces of quarks with the other two out of three colors (red
and green). The gapless CFL phase (gCFL) may prevail over the CFL and
2SC phases at intermediate values of $m^2_s/\mu$ with gaps given
obeying the relation $\Delta_3 > \Delta_2 > \Delta_1 > 0.$ For
chemical potentials that are of astrophysical interest, $\mu <
1000$~MeV, the gap is between 50 and 100~MeV. The order of magnitude
of this result agrees with calculations based on phenomenological
effective interactions \cite{rapp98+99:a,alford99:b} as well as with
perturbative calculations for $\mu > 10$~GeV \cite{son99:a}. We also
note that superconductivity modifies the equation of state at the
order of $(\Delta / \mu)^2$ \cite{alford03:b,alford04:a}, which is
even for such large gaps only a few percent of the bulk energy. Such
small effects may be safely neglected in present determinations of
models for the equation of state of quark-hybrid stars. There has been
much recent work on how color superconductivity in neutron stars could
affect their properties
\cite{rajagopal01:a,alford01:a,alford00:a,rajagopal00:a,alford00:b,%
blaschke99:a}.  These studies reveal that possible signatures include
the cooling by neutrino emission, the pattern of the arrival times of
supernova neutrinos, the evolution of neutron star magnetic fields,
rotational stellar instabilities, and glitches in rotation
frequencies.

\goodbreak
\section{Strange Quark Matter}\label{sec:sqm}

It is most intriguing that for strange quark matter made of more than
a few hundred up, down, and strange quarks, the energy of strange
quark matter may be well below the energy of nuclear matter
\cite{bodmer71:a,witten84:a,terazawa89:a}, $E/A=
930$~MeV, which gives rise to new and novel classes of strange matter
objects, ranging from strangelets at the low baryon-number end to
strange stars at the high baryon number end
\cite{weber99:book,weber05:a,alcock86:a,madsen98:b,alford06:a}.
A simple estimate indicates that for strange quark matter $E/A = 4 B
\pi^2/ \mu^3$, so that bag constants of $B=57~\mevt$ (i.e.
$\bag=145$~MeV) and $B=85~\mevt$ ($\bag=160$~MeV) would place the
energy per baryon of such matter at $E/A=829$~MeV and 915~MeV,
respectively, which correspond obviously to strange quark matter which
is absolutely bound with respect to nuclear
matter \cite{madsen98:b,madsen88:a}.

\subsection{Nuclear crust on strange stars}\label{sec:ncrust}

Strange quark matter is expected to be a color superconductor which,
at extremely high densities, should be in the CFL
phase \cite{rajagopal01:a,alford01:a}. This phase is rigorously
electrically neutral with no electrons
required \cite{rajagopal01:b}. For sufficiently large strange quark
masses, however, the low density regime of strange quark matter is
rather expected to form other condensation patterns (e.g.\ 2SC,
CFL-$K^0$, CFL-$K^+$, CFL-$\pi^{0,-}$) in which electrons are
present \cite{rajagopal01:a,alford01:a}. The presence of electrons
causes the formation of an electric dipole layer on the surface of
strange matter, with huge electric fields on the order of
$10^{19}$~V/cm, which enables strange quark matter stars to be
enveloped in nuclear crusts made of ordinary atomic
\begin{figure}[tb]
\centerline{\psfig{figure=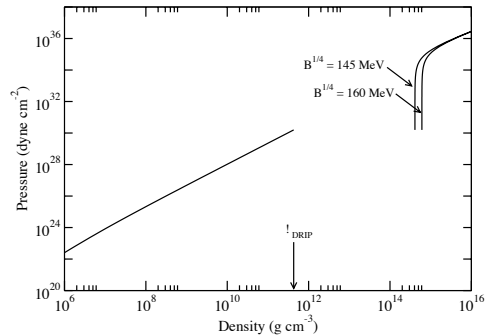,width=7.0cm}}
\caption[]{Illustration of the \eos of strange stars with nuclear
crusts (from \cite{mathews06:a}).}
\label{fig:wdeos} 
\end{figure} 
matter \cite{alcock86:a,alcock88:a,stejner05:a,kettner94:b}.\footnote{Depending
on the surface tension of blobs of strange matter and screening
effects, a heterogeneous crust comprised of blobs of strange quark
matter embedded in an uniform electron background may exist in the
surface region of strange stars \cite{jaikumar05:a}. This heterogeneous
strange star surface would have a negligible electric field which
would make the existence of an ordinary nuclear crust, which requires
a very strong electric field, impossible.}  The maximal possible
density at the base of the crust (inner crust density) is determined
by neutron drip, which occurs at about $4\times 10^{11}~\gcmt$ or
somewhat below \cite{stejner05:a}. The \eos of such a system is shown
in Fig.\ \ref{fig:wdeos}. Sequences of compact strange stars with and
without (bare) nuclear crusts are shown in Fig.\ \ref{fig:MvsR}. Since
the nuclear crust is gravitationally bound to the quark matter core, the
mass-radius relationship of strange stars with crusts resembles the
one of neutron stars and even that of white
dwarfs \cite{glen94:a}. Bare strange stars obey $M \propto R^3$ because
the mass density of quark matter is almost constant inside strange
stars.

\subsection{Strange dwarfs}

For many years only rather vague tests of the theoretical mass-radius
relationship of white dwarfs were possible. Recently the quality and
quantity of observational data on the mass-radius relation of white
dwarfs has been reanalyzed and profoundly improved by the availability
of Hipparcos parallax measurements of several white
dwarfs \cite{provencal98:a}. In that work Hipparcos parallaxes were
used to deduce luminosity radii for 10 white dwarfs in visual binaries
of common proper-motion systems as well as 11 field white
dwarfs. Complementary HST observations have been made to better
determine the spectroscopy for Procyon~B \cite{provencal02:a} and
pulsation of G226-29 \cite{kepler00:a}.  Procyon~B at first appeared as
a rather compact star which, however, was later confirmed to lie on
\begin{figure}[tb]
\centerline{\psfig{figure=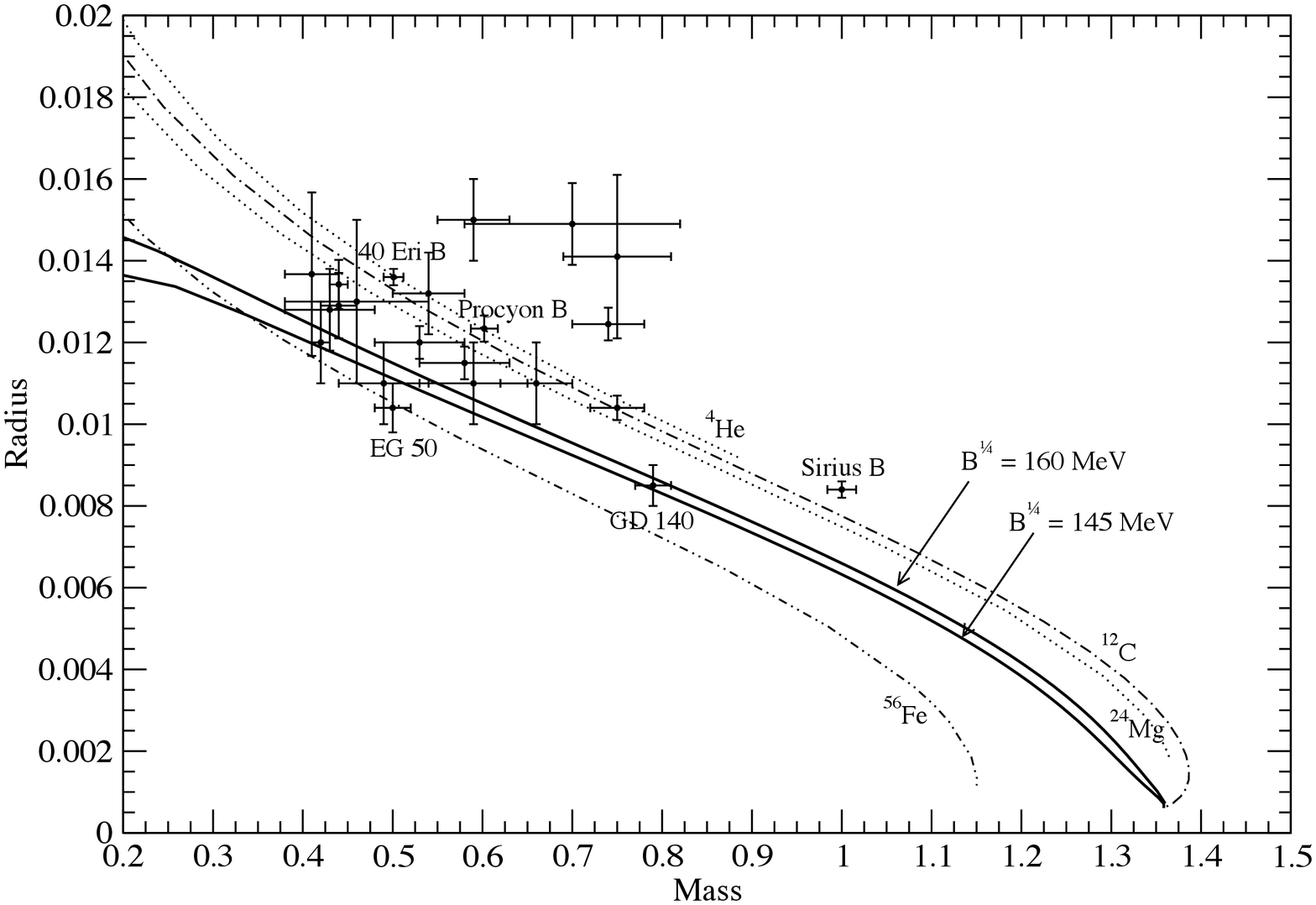,width=10.0cm}}
\caption[]{Comparison of the theoretical mass-radius relationships of
strange dwarfs (solid curves) and normal white
dwarfs \cite{mathews06:a}. Radius and mass are in units of $R_\odot$
and $\msun$, respectively.}
\label{fig:sequences}
\end{figure}
the normal mass-radius relation of white dwarfs.  Stars like Sirius~B
and 40~Erin~B, fall nicely on the expected mass-radius relation too.
Several other stars of this sample (e.g. GD~140, G156--64, EG~21,
EG~50, G181--B5B, GD~279, WD2007--303, G238--44) however appear to be
unusually compact and thus could be strange dwarf candidates
 \cite{mathews04:a}.  The situation is graphically summarized in
Fig.\ \ref{fig:sequences}.

\subsection{Surface properties of strange matter}\label{sec:spss}

The electrons surrounding strange quark matter are held to quark
matter electrostatically. Since neither component, electrons and quark
matter, is held in place gravitationally, the Eddington limit to the
luminosity that a static surface may emit does not apply, and thus the
object may have photon luminosities much greater than $10^{38}~\ergs$.
It was shown by Usov \cite{usov98:a} that this value may be exceeded by
many orders of magnitude by the luminosity of $e^+ e^-$ pairs produced
by the Coulomb barrier at the surface of a hot strange star. For a
surface temperature of $\sim 10^{11}$~K, the luminosity in the
outflowing pair plasma was calculated to be as high as $\sim 3 \times
10^{51}~\ergs$.  Such an effect may be a good observational signature
of bare strange stars \cite{usov98:a,usov01:c,usov01:b,cheng03:a}. If
the strange star is enveloped by a nuclear crust however, which is
gravitationally bound to the strange star, the surface made up of
ordinary atomic matter would be subject to the Eddington limit. Hence
the photon emissivity of such a strange star would be the same as for
an ordinary neutron star.  If quark matter at the stellar surface is
in the CFL phase the process of $e^+ e^-$ pair creation at the stellar
quark matter surface may be turned off, since cold CFL quark matter is
electrically neutral so that no electrons are required and none are
admitted inside CFL quark matter \cite{rajagopal01:b}. This may be
different for the early stages of a hot CFL quark star \cite{vogt03:a}.

\section{Proto-Neutron Star Matter}\label{sec:pnsm}

Here we take a brief look at the composition of proto-neutron star
matter. The composition is determined by the requirements of charge
neutrality and equilibrium under the weak processes, $B_1 \rightarrow
B_2 + l + \bar\nu_l$ and $B_2 + l \rightarrow B_1 + \nu_l$, where
$B_1$ and $B_2$ are baryons, and $l$ is a lepton, either an electron
or a muon. For standard neutron star matter, where the neutrinos have
left the system, these two requirements imply that $Q = \sum_{i} q_i
n_{B_i} + \sum_{l=e, \mu} q_l n_l = 0$ (electric charge neutrality)
and $\mu_{B_i} = b_i \mu_n - q_i \mu_l$ (chemical equilibrium), where
$q_{i/l}$ denotes the electric charge density of a given particle, and
$n_{B_i}$ ($n_l$) is the baryon (lepton) number density.  The
subscript $i$ runs over all the baryons considered. The symbol
$\mu_{B_i}$
\begin{figure}[tb]
\begin{center}
\parbox[t]{5.7cm} {\psfig{file=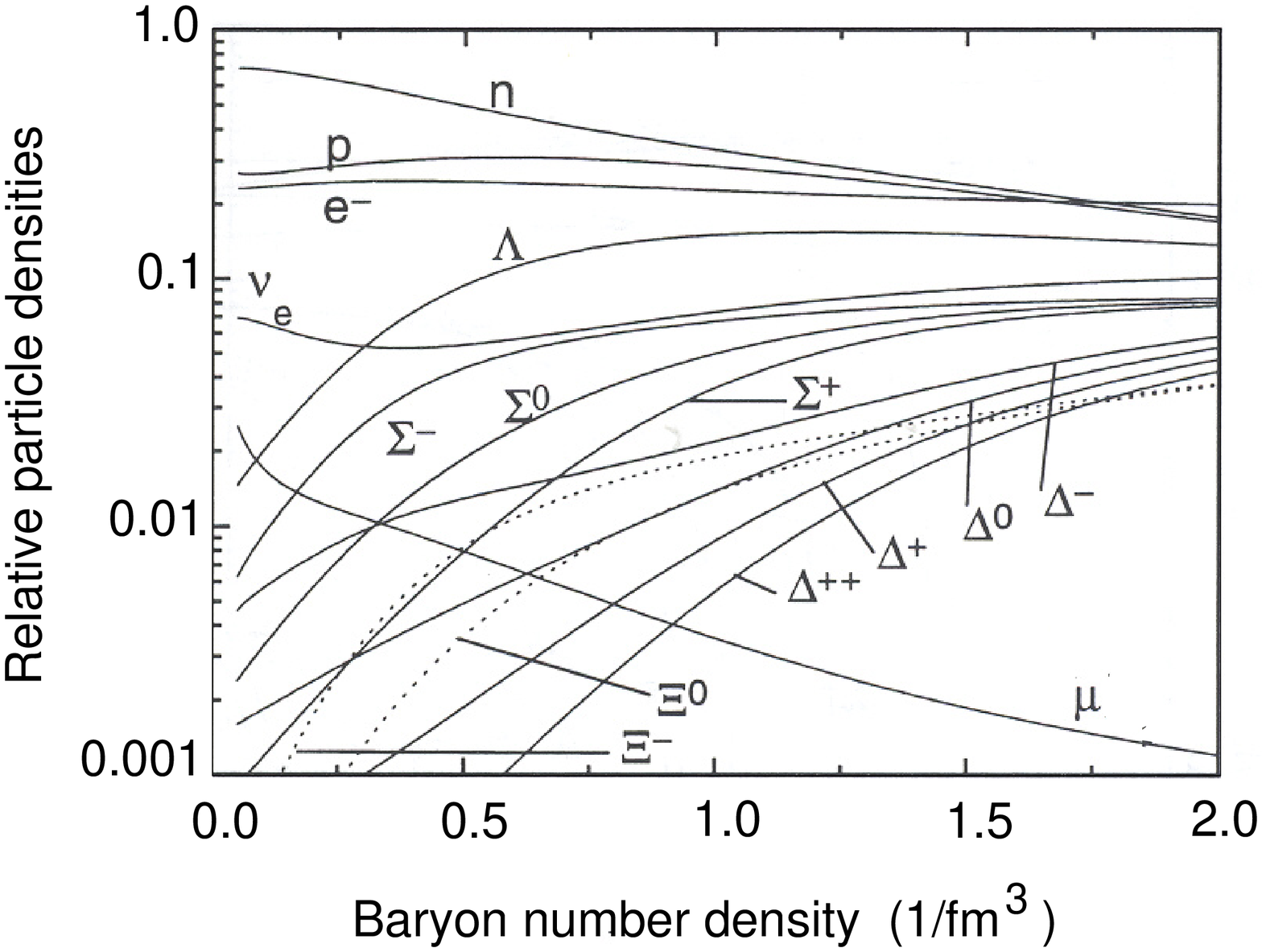,width=5.7cm}
{\caption[]{Composition of hot ($T=40$~MeV) proto-neutron star matter
for $Y_L = 0.3$ \cite{weber06:iship}.}
\label{fig:hotpnsm}}}
\ \hskip 0.1cm \
\parbox[t]{5.7cm}
{\psfig{file=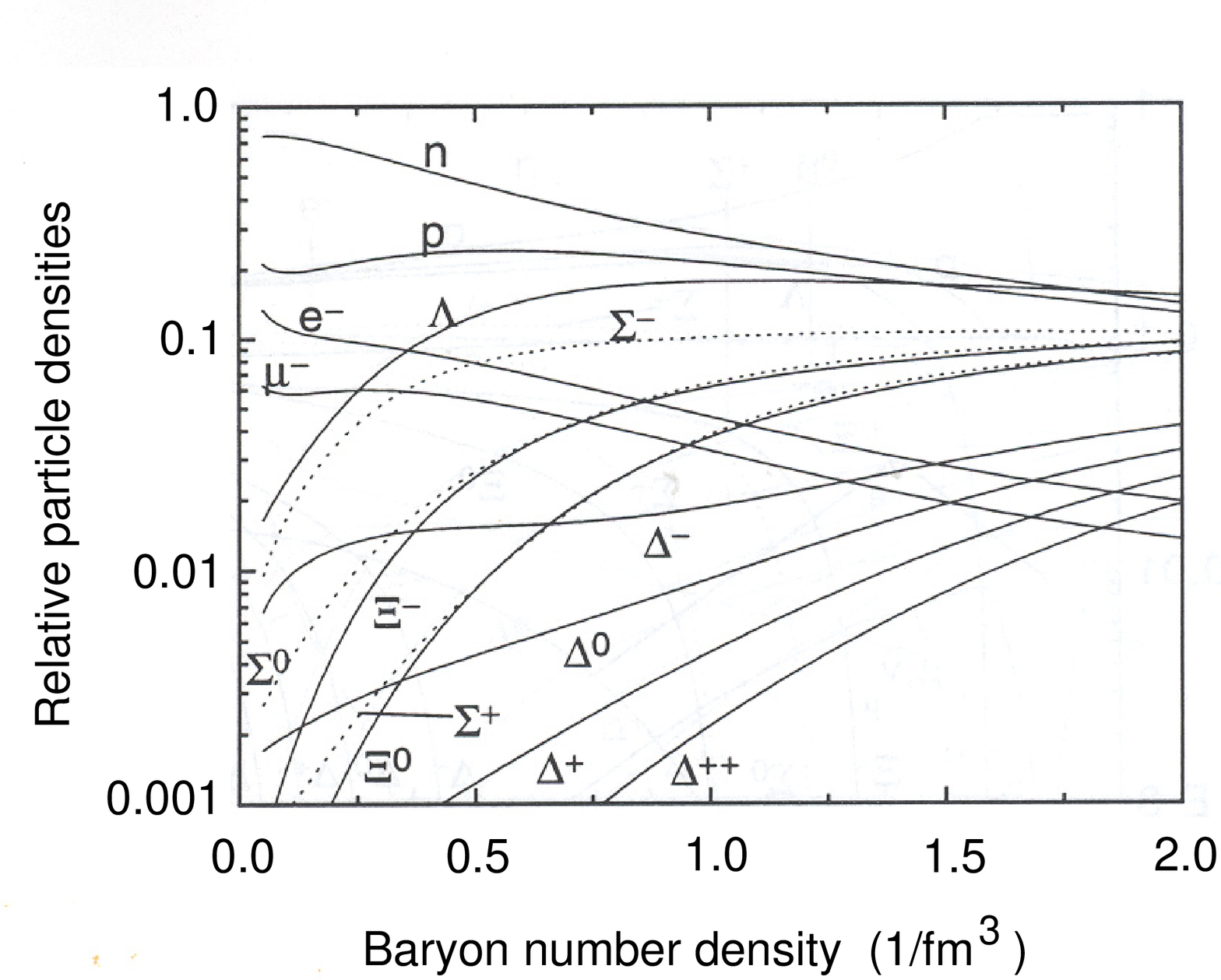,width=5.7cm} {\caption[]{
Same as Fig.\ \ref{fig:hotpnsm}, but for standard neutron star
matter \cite{weber06:iship}.}
\label{fig:hotnsm}}}
\end{center}
\end{figure}
refers to the chemical potential of baryon $i$, $b_i$ is the
particle's baryon number, and $q_i$ is its charge. The chemical
potential of the neutron is denoted by $\mu_n$.  When the neutrinos
are trapped, as it is the case for proto-neutron star matter, the
chemical equilibrium condition is altered to $\mu_{B_i} = b_i \mu_n -
q_i (\mu_l - \mu_{\nu_l})$ and $\mu_e - \mu_{\nu_e} = \mu_\mu -
\mu_{\nu_\mu}$, where $\mu_{\nu_l}$ is the chemical potential of the
neutrino $\nu_l$. In proto-neutron star matter, the electron lepton
number $Y_L = (n_e+n_{\nu_e})/n_B$ is initially fixed at a value of
around $Y_{L_e} = Y_e + Y_{\nu_e} \simeq 0.3 - 0.4$ as suggested by
gravitational collapse calculations of massive stars. Also, because no
muons are present when neutrinos are trapped, the constraint
$Y_{L_\mu} = Y_\mu + Y_{\nu_mu} =0$ can be imposed. Figures
\ref{fig:hotpnsm} and \ref{fig:hotnsm} show sample compositions of
proto-neutron star matter and standard neutron star matter (no
neutrinos) computed for the relativistic mean-field approximation. The
presence of the $\Delta$ particle in (proto) neutron star matter at
finite temperature is striking. This particle is generally absent in
cold neutron star matter treated in the relativistic mean-field
approximation \cite{glen97:book,weber99:book,glen85:a}.

\goodbreak
\section{Rotational Instabilities}\label{sec:r-mode}

\begin{figure}[tb]
\centerline{\psfig{figure=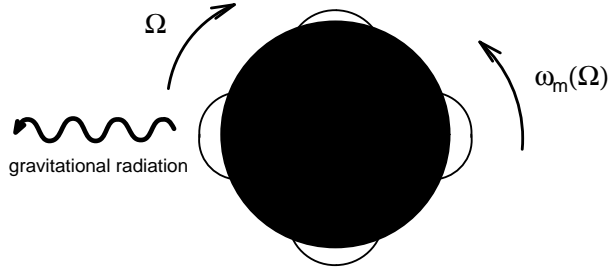,width=8.0cm,angle=90}}
\caption[]{Representation of an $m=4$ perturbation of a rotating neutron star.
  $\Omega$ denotes the star's rotational frequency, $\omega_m$ is the
  frequency of the counter-rotating perturbation \cite{weber99:book}.}
\label{fig:GRR}
\end{figure}

An absolute limit on rapid rotation is set by the onset of mass shedding from
the equator of a rotating star.  However, rotational instabilities in rotating
stars, known as gravitational radiation driven instabilities, set a more
stringent limit on rapid stellar rotation than mass shedding.  These
instabilities originate from counter-rotating surface vibrational modes which
at sufficiently high rotational star frequencies are dragged forward, as
schematically illustrated in Fig.\ \ref{fig:GRR}.
In this case gravitational radiation, which inevitably accompanies the
aspherical transport of matter, does not damp the instability modes but rather
drives them. Viscosity plays the important role of damping these instabilities
at a sufficiently reduced rotational frequency such that the viscous damping
rate and power in gravity waves are comparable.
\begin{figure}[tb]
\begin{center} 
\parbox[t]{5.7cm} 
{\psfig{figure=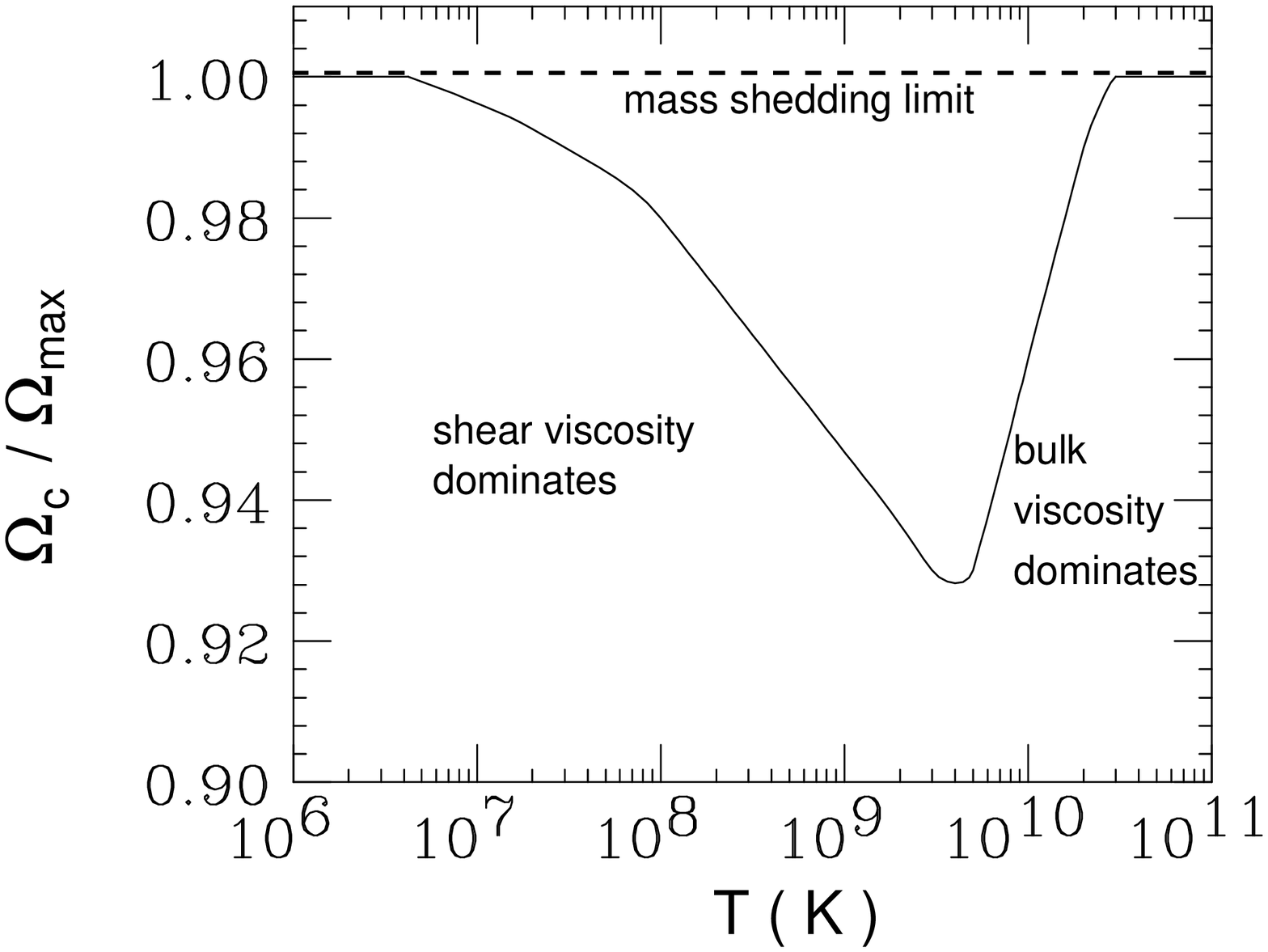,width=5.5cm,angle=0} 
{\caption[]{Gravitational radiation driven f-mode instability suppressed by
shear and bulk viscosity. (Fig.\ from  \cite{weber99:book}.)}
\label{fig:OgrrvsTf}}}
\ \hskip 0.1cm \
\parbox[t]{5.7cm}
{\psfig{figure=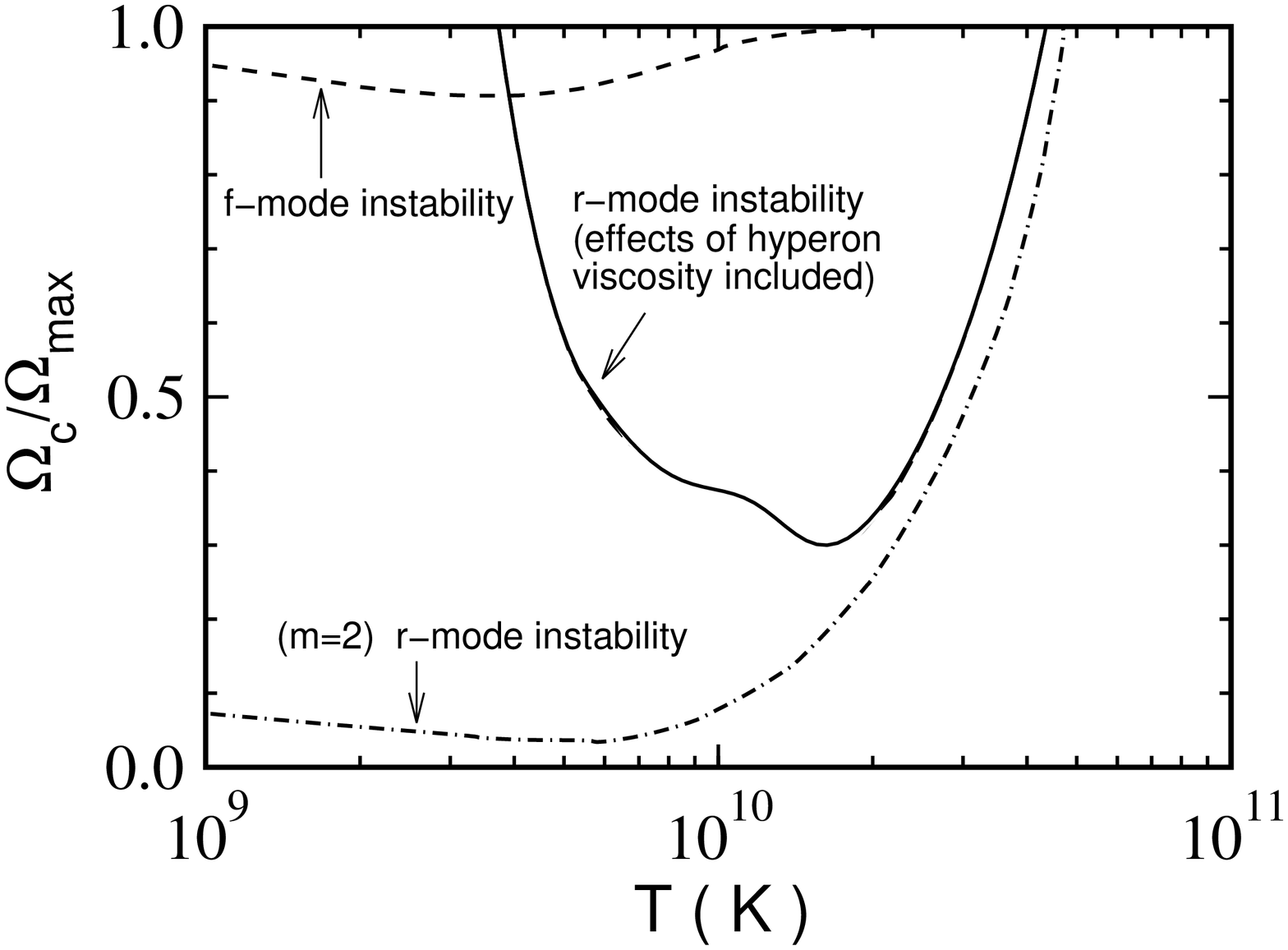,width=5.5cm,angle=0} 
{\caption[]{Comparison of f-mode instability with r-mode instability.
(Data from Refs.\  \cite{lindblom01:proc,lindblom02:a}.)}
\label{fig:OgrrvsTr}}}
\end{center}
\end{figure}
The most critical instability modes that are driven unstable by gravitational
radiation are f-modes \cite{weber99:book,lindblom01:proc} and r-modes
\cite{andersson98:a,andersson01:a}.  Figure \ref{fig:OgrrvsTf} shows the
stable neutron star frequencies if only f-modes were operative in neutron
star. One sees that hot as well as cold neutron stars can rotate at
frequencies close to mass shedding, because of the large contributions of
shear and bulk viscosity, respectively, for this temperature regime.  The more
recently discovered r-mode instability \cite{andersson98:a,friedman98:a} may
change the picture completely, as can be seen from Fig.\ \ref{fig:OgrrvsTr}.
These modes are driven unstable by gravitational radiation over a considerably
wider range of angular velocities than the f-modes (cf.\ dashed curve labeled
($m=2$) r-mode instability). In stars with cores cooler than $\sim 10^9$~K, on
the other hand, the r-mode instability may be completely suppressed by the
viscosity originating from the presence of hyperons in neutron star matter, so
that stable rotation would be limited by the f-mode instability again
\cite{lindblom02:a}.

\begin{figure}[tb]
\begin{center} 
\parbox[t]{5.7cm} {\psfig{file=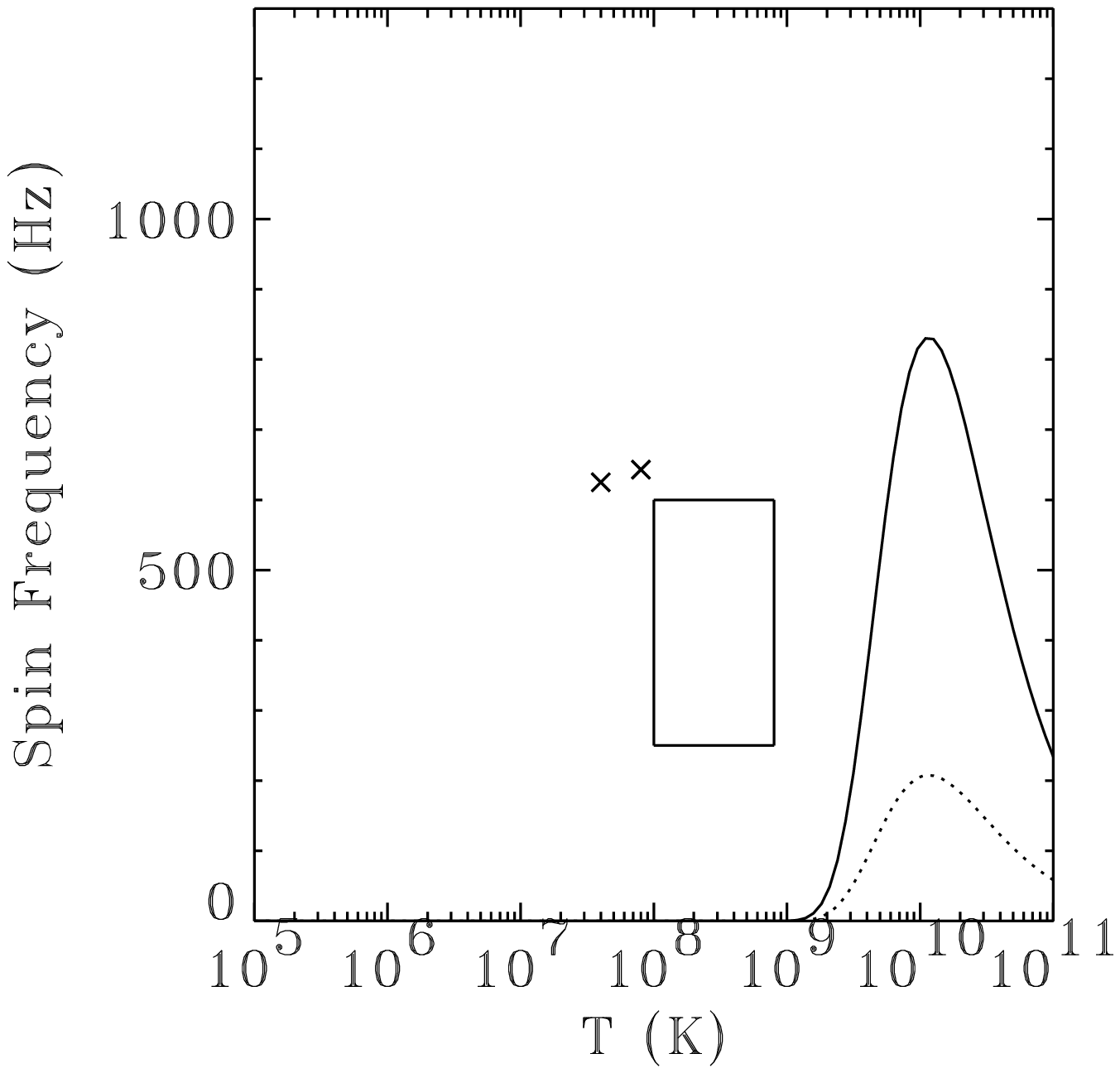,width=5.5cm}
{\caption{Critical rotation frequencies versus stellar temperature for
CFL strange stars \cite{madsen00:b}.}
\label{fig:cfl}}}
\ \hskip .1 cm \ 
\parbox[t]{5.7cm} {\psfig{file=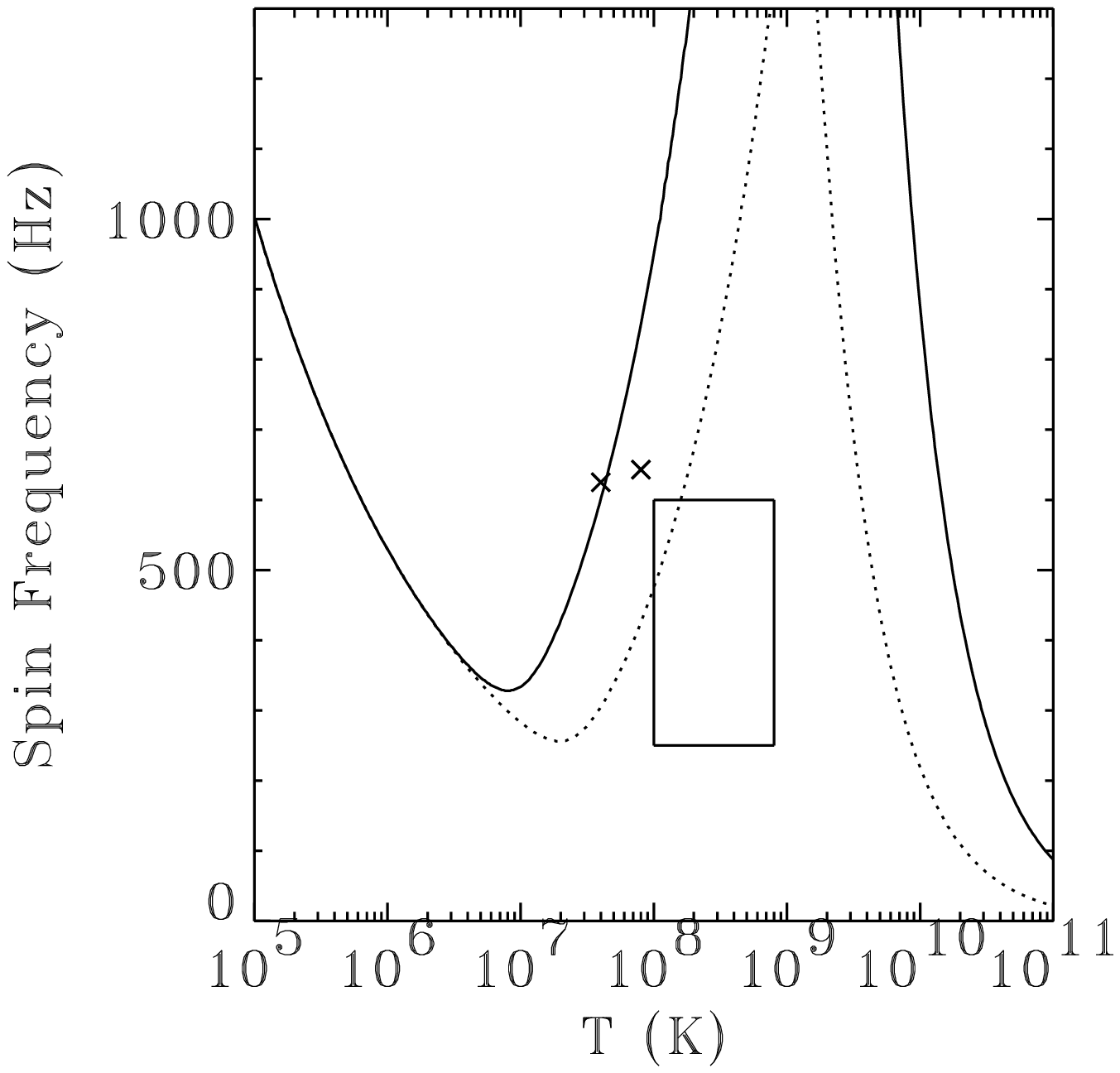,width=5.5cm}
{\caption{Same as Fig.\ \ref{fig:cfl}, but for 2SC quark
stars \cite{madsen00:b}.}
\label{fig:2sc}}}
\end{center}
\end{figure}

Figures \ref{fig:cfl} and \ref{fig:2sc} are the counterparts to Figs.\
\ref{fig:OgrrvsTf} and \ref{fig:OgrrvsTr} 
but calculated for strange stars made of CFL and 2SC quark matter,
respectively \cite{madsen98:a,madsen00:b}. The r-mode instability
seems to rule out that pulsars are CFL strange stars, if the
characteristic time scale for viscous damping of r-modes are
exponentially increased by factors of $\sim \Delta/T$ as calculated in
\cite{madsen98:a}. An energy gap as small as $\Delta = 1$~MeV was
assumed. For much larger gaps of $\Delta \sim 100$~MeV, as expected
for color superconducting quark matter (see section \ref{sec:color}),
the entire diagram would be r-mode unstable.  The full curve in
Fig.\ \ref{fig:cfl} is calculated for a strange quark mass of $m_s =
200$~MeV, the dotted curve for $m_s=100$~MeV.  The box marks the
positions of most low mass X-ray binaries (LMXBs) \cite{klis00:a}, and
the crosses denote the most rapidly rotating millisecond pulsars
known. All strange stars above the curves would
spin down on a time scale of hours due to the r-mode instability, in
complete contradiction to the observation of millisecond pulsars and
LMXBs, which would rule out CFL quark matter in strange stars (see,
however, \cite{manuel04:a}). Figure \ref{fig:2sc} shows the
critical rotation frequencies of quark stars as a function of internal
stellar temperature for 2SC quark stars.  For such quark stars the
situation is less conclusive.  Rapid spin-down, driven by the r-mode
gravitational radiation instability, would happen for stars above the
curves.

\goodbreak
\section{Net Electric Fields and Compact Star Structure}

Here we consider the possibility that the electric charge density
inside compact stars (neutron stars, strange stars) is not identically
zero.  This may be the case, for example, for compact stars accreting
ionized hydrogen. Another example are strange quark stars. They could
have electric charge distributions on their surfaces that generate
electric fields on the order of $10^{18}$~V/cm
\cite{weber99:book,weber05:a,alcock86:a,kettner94:b} for ordinary
quark matter, and $10^{19}$~V/cm \cite{usov04:a} if quark matter is a
color-superconductor. Although the electric field on strange stars
exists only in a very narrow region of space, it is interesting to
study the effects of such ultra-high electric fields on the structure
of the star.

It has already been shown that the energy densities of ultra-high
electric fields can substantially alter the structure (mass--radius
relationship) of compact stars \cite{ray03:a}, depending on the
strength of the electric field.  In contrast to electrically uncharged
stars, the energy-momentum tensor of charged stars has two key
contributions, the usual matter-energy term plus the energy density
term that originates from the electric field. The latter plays a dual
role for compact star physics. Firstly, it acts as an additional
source of gravity and, secondly, it introduces Coulomb interactions
inside the star. Both features can alter the properties of compact
stars significantly, as we shall demonstrate below.

We will restrict ourselves to spherically symmetric compact stars. The
metric of such objects is given by
\begin{equation}
ds^2=e^{\nu(r)}c^{2}dt^{2}-e^{\lambda(r)}dr^{2}-r^{2}(d\theta^{2}
+ \sin^{2} \theta d\phi^{2})\, . \label{metr}
\end{equation}
The energy-momentum tensor consists of the usual perfect fluid term
supplemented with the electromagnetic energy-momentum tensor,
\begin{eqnarray}
T_{\kappa}{}^{\mu} = (p +\rho c^2)u_{\kappa} u^{\mu} + p\delta_{\kappa}{}^{
 \mu} +\frac{1}{4\pi} \left[ F^{\mu l} F_{\kappa l} +\frac{1}{4
\pi} \delta_{\kappa}{}^{\mu} F_{kl} F^{kl} \right] \, ,
\end{eqnarray}
where $u^{\mu}$ is the fluid's four-velocity, $p$ and $\rho c^2\equiv
\epsilon$ are the pressure and energy density, respectively, and $F^{\mu
\kappa}$ satisfies the covariant Maxwell equation,
\begin{equation}
[(-g)^{1/2} F^{\kappa \mu}]_{, \mu} = 4\pi J^{\kappa} (-g)^{1/2}
\label{ecem}.
\end{equation}
The quantity $J^{\kappa}$ denotes the four-current which represents the
electromagnetic sources in the star. For a static spherically
symmetric system, the only non-zero component of the four-current is
$J^{1}$, which implies that the only non-vanishing component of $F^{\kappa
\mu}$ is $F^{01}$. We therefore obtain from Eq.\ (\ref{ecem})
\begin{equation}
F^{01}(r)= E(r) = r^{-2} \, e^{-(\nu + \lambda)/2} \int_{0}^{r} 4\pi
j^{0} e^{(\nu + \lambda )/2} dr \, , \label{comp01}
\end{equation}
which is nothing other than the electric field. This relation can be
identified as the relativistic version of Gauss' law. In addition we
see that the electric charge of the system is given by
\begin{equation}
Q(r) = \int_{0}^{r} 4\pi j^{0} e^{(\nu + \lambda )/2} dr \, .  \label{Q}
\end{equation}
With the aid of Eq.\ (\ref{Q}) the energy-momentum tensor of the
system can be written as
%\begin{small}
\begin{equation}
T_{\kappa}{}^{\mu} =\left( \begin{array}{cccc}
-\left( \epsilon + \frac{Q^2(r)}{8\pi r^4} \right)  & 0 & 0 & 0 \\ 
0 & p - \frac{Q^2(r)}{8\pi r^4} & 0 & 0 \\ 
0 & 0 & p + \frac{Q^2(r)}{8\pi r^4}  & 0 \\ 
0 & 0 & 0 & p  +\frac{Q^2(r)}{8\pi r^4}
\end{array} \right). \label{TEMch}
\end{equation}
%\end{small}
Using the energy-momentum tensor (\ref{TEMch}), Einstein's field
equation leads to
\begin{eqnarray}
e^{-\lambda}\left(
-\frac{1}{r^{2}}+\frac{1}{r}\frac{d\lambda}{dr}\right)
+\frac{1}{r^{2}}=\frac{8\pi G}{c^4} \left( p - \frac{Q^{2}(r)}{8\pi
r^4} \right), \label{fe1q} \\
e^{-\lambda}\left(\frac{1}{r}\frac{d\kappa}{dr}+\frac{1}{r^{2}}\right)
-\frac{1}{r^{2}}= - \frac{8 \pi G}{c^4} \left( \epsilon +
\frac{Q^{2}(r)}{8\pi r^4} \right)  . \label{fe2q}
\end{eqnarray}
At this point we define the radial component of the metric $g^{11}$,
in analogy to the exterior solution of Reissner-Nordstr\"{o}m, as
\cite{bekenstein71:a}
\begin{equation}
e^{-\lambda}(r) = 1 - \frac{Gm(r)}{rc^2} +\frac{GQ^2(r)}{r^2
c^4} \, . \label{nord}
\end{equation}
From equations (\ref{fe1q}), (\ref{fe2q}) and (\ref{nord}), we derive
an expression for $m(r)$, which is interpreted as the total mass of
the star at a radial distance $r$. This expression reads
\begin{equation}
\frac{dm(r)}{dr} = \frac{4\pi r^2}{c^{2}} \epsilon +\frac{Q(r)}{c^2
r}\frac{dQ(r)}{dr} \, , \label{dmel}
\end{equation}
which reveals that, in addition to the standard term originating from
the eos of the stellar fluid, the electric field energy contributes to
the star's total mass too.  Next, we impose the vanishing of the
divergence of the energy-momentum tensor, $T^{\mu}{}_{\kappa;\mu} =0$,
which leads to the Tolman-Oppenheimer-Volkoff (TOV) equation of
electrically charged stars,
\begin{eqnarray}
\frac{dp}{dr} = - \frac{2G\left[ m(r) +\frac{4\pi r^3}{c^2} \left( p -
\frac{Q^{2} (r)}{4\pi r^{4} c^{2}} \right) \right]}{c^{2} r^{2} \left(
1 - \frac{2Gm(r)}{c^{2} r} + \frac{G Q^{2}(r)}{r^{2} c^{4}} \right)}
(p +\epsilon) +\frac{Q(r)}{4 \pi r^4}\frac{dQ(r)}{dr} \,
. \label{TOVca}
\end{eqnarray}
Summarizing the relevant stellar structure equations, we end up with
the following set of equations:
\begin{eqnarray}
\frac{d\lambda}{dr} &=& \frac{8\pi G}{c^{4}}
\left( \epsilon +\frac{Q^{2}(r)}{8 \pi r^{4}} \right) r 
e^{\lambda} - \left(\frac{e^{-\lambda} - 1}{r} \right) \, ,
\label{nuu} \\
\frac{d\nu}{dr} &=& \frac{2G\left[ m(r) +\frac{4\pi r^3}{c^2}
\left( p - \frac{Q^{2} (r)}{4\pi r^{4} c^{2}} \right) \right]}{c^{2} r^{2}
\left( 1 - \frac{2Gm(r)}{c^{2} r} + \frac{G Q^{2}(r)}{r^{2} 
c^{4}} \right)} \label{dnuu} \, . \\
\frac{dm(r)}{dr} &=& \frac{4\pi r^2}{c^{2}} \epsilon
+\frac{Q(r)}{c^2 r}\frac{dQ(r)}{dr}, \label{dme2} \\
\frac{dQ(r)}{dr} &=& 4 \pi r^{2} j^{0} e^{-(\nu +\lambda) /2 } \, .
\label{dq2}\\
\frac{dp}{dr} &=& - \frac{2G\left[ m(r) +\frac{4\pi r^3}{c^2}
\left( p - \frac{Q^{2} (r)}{4\pi r^{4} c^{2}} \right) \right]}{c^{2} r^{2}
\left( 1 - \frac{2Gm(r)}{c^{2} r} + \frac{G Q^{2}(r)}{r^{2} c^{4}} \right)}
(p +\epsilon) +\frac{Q(r)}{4 \pi r^4}\frac{dQ(r)}{dr}
 \label{TOV2} \, .
\end{eqnarray}
Equations (\ref{nuu}) and (\ref{dnuu}) arise from Einstein's field equation,
Eq.\ (\ref{dme2}) is the mass continuity equation, Eq.\ (\ref{dq2})
comes from the Maxwell equations, and Eq.\ (\ref{TOV2}) is the TOV
equation.  This system of coupled differential equations is
subject to the following boundary conditions
\begin{eqnarray}
p(0)=p_c \, , \quad e^{\lambda}=0 \, , 
Q(0)=0\, , \quad  m(0)=0 \, .
\end{eqnarray}
In addition to these conditions, one needs to specify the star's
central density (or, equivalently, the central pressure) for a given
equation of state and a given electric charge distribution. This will
be discussed in more details in the next sections.

As already mentioned at the beginning of this section, strange stars
may be expected to carry huge electric fields on their surfaces
\cite{weber99:book,weber05:a,alcock86:a,kettner94:b,usov04:a}. We want to
study the effects of such fields on the overall structure of strange
stars. To this aim, we model the charge distribution by superimposing
two Gaussian functions.
\begin{figure}
\centerline{\psfig{file=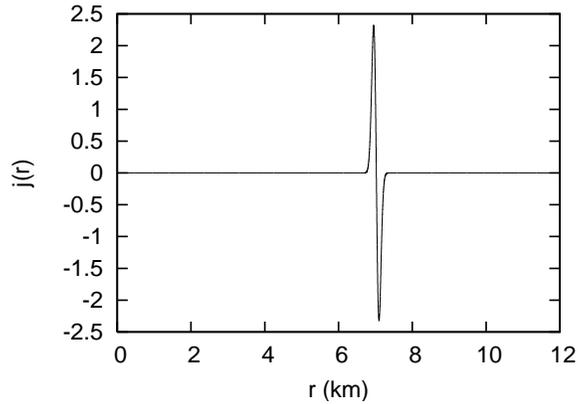,width=8cm,clip=}}
\caption{Displacement of electric charges on the surface of a strange star.
The mathematical form is obtained by superimposing two Gaussian
functions.}
\label{chdist2}
\end{figure}
The first Gaussian is chosen to be positive, representing the
accumulation of a net positive charge. The second Gaussian, slightly
displaced from the first one, is chosen negative to represent the
accumulation of a net negative charge. Mathematically, we thus have
\begin{equation}
j(r) = \frac{\sigma}{b\sqrt{\pi}} \left( e^{- \left( \frac{r-r_1}{b}
\right)^2} -e^{- \left( \frac{r-r_2}{b} \right)^2} \right) ,
\label{eq:gauss}
\end{equation}
where $\sigma$ is a constant that controls the magnitude of the
Gaussians and $b$ the widths of the Gaussians. The graphical
illustration of Eq.\ (\ref{eq:gauss}) is shown in Fig.\
\ref{chdist2}. To obtain a noticeable impact of the electric field on
the structure of strange stars, one needs to have Gaussians with a
width of at least around $0.05$~km. For such widths we find the
\begin{figure}
\centerline{\psfig{file=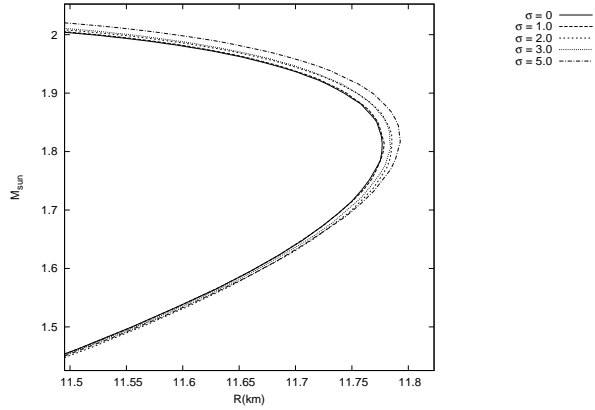,width=8cm,clip=}}
\caption{Mass--radius relationships of electrically charged strange stars.}
\label{chdist1}
\end{figure}
mass--radius relationships shown in Fig.\ \ref{chdist1}. The
deviations from the mass-radius relationships of uncharged strange
\begin{figure}
\centerline{\psfig{file=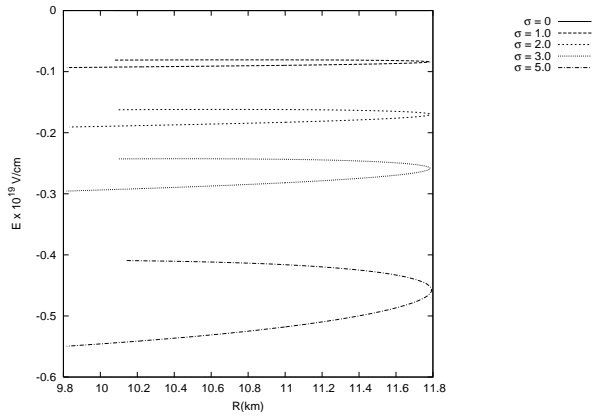,width=8cm,clip=} }
\caption{Electric fields at the surface of strange stars.}
\label{ExR}
\end{figure}
stars are found to increase with mass, and are largest for the
maximum-mass star of each stellar sequence.

The radial distribution of the electric charge over the surface of a
strange star is particularly interesting. The reason being the
occurrence of the metric functions in Eq.\ (\ref{Q}), which defines
the star's total net charge. Since the metric functions are not
symmetric in the radial distance, the charge distribution is rendered
asymmetric and stars that are strictly electrically charge neutral in
flat space-time become electrically charged and thus possess non-zero
electric fields. Figures \ref{ExR} shows the electric field at the
surface of strange stars.
\begin{figure}
\centerline{\psfig{file=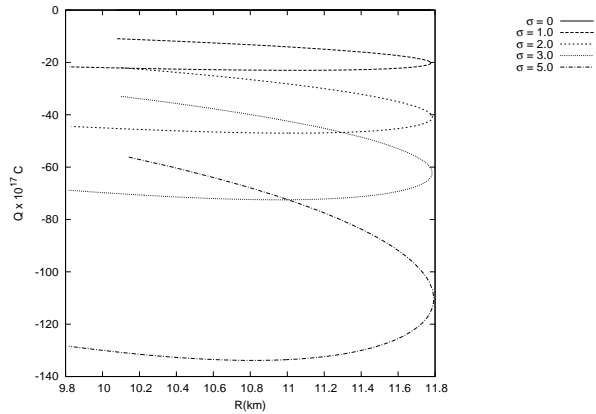,width=8cm,clip=}}
\caption{Electric charge on strange quark stars.}
\label{QxR} 
\end{figure}
Figure \ref{QxR} shows the net electric charge at the surface of
strange stars. Both plots account for the general relativistic charge
separation effect.

\goodbreak
\section{Conclusions and Outlook}

It is often stressed that there has never been a more exciting time in the
overlapping areas of nuclear physics, particle physics and relativistic
astrophysics than today.  This comes at a time where new orbiting
observatories such as the Hubble Space Telescope (HST), Rossi X-ray Timing
Explorer, Chandra X-ray satellite, and the X-ray Multi Mirror Mission (XMM
Newton) have extended our vision tremendously, allowing us to observe compact
star phenomena with an unprecedented clarity and angular resolution that
previously were only imagined.  On the Earth, radio telescopes like Arecibo,
Green Bank, Parkes, VLA, and instruments using adaptive optics and other
revolutionary techniques have exceeded previous expectations of what can be
accomplished from the ground. Finally, the gravitational wave detectors LIGO,
LISA, VIRGO, and Geo-600 are opening up a window for the detection of
gravitational waves emitted from compact stellar objects such as neutron stars
and black holes. This unprecedented situation is providing us with key
information on neutron stars, which contain cold and ultra-dense baryonic
matter permanently in their cores.  As discussed in this paper, a key role in
neutron star physics is played by strangeness. It alters the masses, radii,
moment of inertia, frame dragging of local inertial frames, cooling behavior,
and surface composition of neutron stars.  Other important observables
influenced by strangeness may be the spin evolution of isolated neutron stars
and
\begin{table}[tbh]
\begin{center}
%\begin{minipage}[t]{14.5 cm}
\caption{Past, present, and future search experiments for strange quark
matter \cite{weber05:a}.}\label{tab:labexp}
%\end{minipage}
\begin{tabular}{l|l} \hline
Experiment                                 &References                                                \\ \hline
Cosmic ray searches for strange nuggets:   &                                                          \\
~~~~~ AMS-02$^{\rm a}$                     & \cite{ams01:homepage,sandweiss04:a}                       \\
~~~~~ CRASH$^{\rm b}$                      & \cite{saito90:a,saito94:a,ichimura93:a}                   \\
~~~~~ ECCO$^{\rm c}$                       & \cite{ecco01:homepage}                                    \\
~~~~~ HADRON                               & \cite{shaulov96:a}                                        \\
~~~~~ IMB$^{\rm d}$                        & \cite{rujula83:a}                                         \\
~~~~~ JACEE$^{\rm e}$                      & \cite{miyamura95:a,lord95:a}                              \\
~~~~~ MACRO$^{\rm f}$                      & \cite{macro92:a,ambrosio00:a,ambrosio02:a,giacomelli02:a} \\
Search for strangelets in terrestrial matter: & \cite{lu04:a}                                          \\
~~~~~ Tracks in ancient mica                     & \cite{rujula84:a,price84:a}                         \\ 
~~~~~ Rutherford backscattering                  & \cite{bruegger89:a,isaac98:a}                       \\
Search for strangelets at accelerators:    &                                                          \\
~~~~~ Strangelet searches E858, E864, E878, E882-B,  & \cite{thomas95:a,rusek96:a,buren99:a}           \\
~~~~~~E896-A, E886                         &                \\
~~~~~ H-dibaryon search                    & \cite{belz96:a,belz96:b}                                               \\
~~~~~ Pb+Pb collisions                     & \cite{dittus95:a,appelquist96:a,ambrosini96:a,klingenberg99:topr}      \\
\hline
\end{tabular}
\begin{minipage}[t]{12.5 cm}
\vskip 0.1cm
\noindent
$^{\rm a}$ AMS: Alpha Magnetic Spectrometer (scheduled for 2005-2008).  \\
$^{\rm b}$ CRASH: Cosmic Ray And Strange Hadronic matter.                \\
$^{\rm c}$ ECCO: Extremely-heavy Cosmic-ray Composition Observer.        \\
$^{\rm d}$ IMB: Irvine Michigan Brookhaven proton-decay detector (1980-1991). \\
$^{\rm e}$ JACEE: Japanese-American Cooperative Emulsion Chamber Experiment.  \\
$^{\rm f}$ MACRO: Monopole, Astrophysics and Cosmic Ray Observatory (1989-2000). 
\end{minipage} 
\vskip -0.4cm
\end{center}
\end{table}
neutron stars in low-mass x-ray binaries. All told, these observables play a
key role for the exploration of the phase diagram of dense nuclear matter at
high baryon number density but low temperature \cite{klahn06:a_short}, which
is not accessible to relativistic heavy ion collision experiments.

Obviously, our understanding of neutron stars has changed dramatically since
their first discovery some 40 years ago. In what follows, I briefly summarize
what we have learned about the internal structure of these fascinating object
since their discovery. I will address some of the most important open
questions regarding the composition of neutron star matter and its associated
equation of state, and will mention new tools, telescopes, observations, and
calculations that are needed to answer these questions:

\begin{itemize}
\item There is no clear picture yet as to what kind of matter exists in the
  cores of neutron stars. They may contain significant hyperon populations,
  boson condensates, a mixed phase of quarks and hadrons, and/or pure quark
  matter made of unconfined up, down, and strange quarks.
\item Pure neutron matter constitutes an excited state relative to many-baryon
  matter and, therefore, will quickly transform via weak reactions to such
  matter.
\item Neutron stars made up of pure, interacting neutron matter cannot rotate
  as rapidly as the very recently discovered pulsars PSR~J1748-2446ad, which
  spins at 716~Hz.  The equation of state of such matter, therfore, imposes an
  upper bound on the equation of state of neutron star matter that is tighter
  than the usual $P=\epsilon$ constraint (see Fig.\ \ref{fig:nucl_eoss}).
\item Charm quarks do not play a role for neutron star physics, since they
  become populated at densities which are around 100 times greater than the
  densities encountered in the cores of neutron stars. While hydrostatically
  stable, ``charm'' stars are unstable against radial oscillations and, thus,
  cannot exist stably in the universe \cite{kettner94:b}.
\item Multi-quark states like the H-particle appear to make 
neutron stars unstable.
\item Significant populations of $\Delta$'s are predicted by relativistic
  Brueckner-Hartree-Fock calculations, but not by standard mean-field
  calculations which do not account for dynamical correlations among baryons
  computed from the relativistic T-matrix equation.
\item The finite temperatures of proto neutron stars favors the population of
  $\Delta$'s already at the mean-field level.
\item The r-modes are of key interest for several reasons: 1.\ they may
  explain why young neutron stars spin slowly, 2.\ why rapidly accreting
  neutron stars (LMXB) spin slowly and within a narrow band, and 3.\ they may
  produce gravitational waves detectable by LIGO. Knowing the bulk viscosity
  originating from processes like $n+n -> p^+ + \Sigma^-$ and the superfluid
  critical temperature of $\Sigma^-$, both are poorly understood at present,
  will be key.
\item The loss of pressure resulting from the appearance of additional
  hadronic degrees of freedom at high densities reduces the (maximum) mass of
  neutron stars. This feature may serve as a key criteria to distinguish
  between, and eliminate certain, classes of equations of state
  \cite{glen97:book,weber99:book,lattimer05:a}.
\item Heavy neutron stars, with masses of around two solar masses, do not
  automatically rule out the presence of hyperons or quarks in the cores of
  neutron stars \cite{alford06:b}.
\item Depending on the densities reached in the cores of neutron stars, both
  Schroedinger-based models as well as relativistic field-theoretical models
  may be applicable to neutron star studies.
\item The density dependence of the coupling constants of particles in
  ultra-dense neutron star matter needs be taken into account in stellar
  structure calculations. Density dependent relativistic field theories are
  being developed which account for this feature..
%\item The same is true for the in-medium behavior of particle masses.
\item The models used to study the quark-hadron phase transition in the cores
  of neutron stars are extremely phenomenological and require considerable
  improvements.
% \item Do neutron stars with masses greater that around two solar masses exist? 
\item If quark matter exists in the cores of neutron stars, it will be a color
  superconductor whose complex condensation pattern is likely to change with
  density inside the star. The exploration of the numerous astrophysical
  facets of (color superconducting) quark matter is therefore of uppermost
  importance.  What are the signatures of color superconducting quark matter
  in neutron stars? So far it has mostly been demonstrated that color
  superconductivity is compatible with observed neutron star properties.
\item A two-step quark-hadron phase transition (1.\ from nuclear matter to
  regular quark matter, 2.\ from regular quark matter to color superconducting
  quark matter) may explain long quiescent gamma-ray bursts due to the two
  phase transitions involved.
\item Are there isolated pulsar that are spinning up? Such a (backbending)
  phenomenon could be caused by a strong first-order-like quark-hadron phase
  transitions in the core of a neutron star
  \cite{glen97:a,zdunik04:a,chubarian00:a}.
\item Was the mass of the neutron star created in SN~1987A around $1.5 \msun$?
  And did SN~1987A go into a black hole or not? If the answer to both
  questions were yes, a serious conflict with the observation of heavy neutron
  stars would arise. On the other hand, it could also indicate the existence
  generically different classes of ``neutron'' stars with very different
  maximum masses.
\item Sources known to increase the masses of neutron stars are differential
  rotation, magnetic fields, and electric fields. Some of these sources are
  more effective (and plausible) than others though.
\item Nuclear processes in non-equilibrium nuclear crusts (e.g. pycnonuclear
  reactions) and/or cores (heating caused by changes in the composition) of
  neutron stars can alter the thermal evolution of such stars significantly.
  We are just beginning to study these processes in greater detail.
\item What is the shell structure for very neutron rich nuclei in the crusts
  of neutron stars?
\item Do N=50 and N=82 remain magic numbers? Such questions will be addressed
  at GSI (Darmstadt) and RIKEN.

\item Are there pulsars that rotate below one millisecond? Such objects may be
  composed of absolutely stable strange quark matter instead of purely
  gravitationally bound hadronic matter.  Experimental physicists have
  searched unsuccessfully for stable or quasistable strange matter systems
  over the past two decades. These searches fall in three main categories: (a)
  searches for strange matter (strange nuggets or strangelets) in cosmic rays,
  (b) searches for strange matter in samples of ordinary matter, and (c)
  attempts to produce strange matter at accelerators.  An overview of these
  search experiments is given in table~\ref{tab:labexp}.
\item Strange stars may be enveloped in a crust. There is a critical surface
  tension below which the quark star surfaces will fragment into a crystalline
  crust made of charged strangelets immersed in an electron gas
  \cite{alford06:a,jaikumar05:a}

\item If bare, the quark star surface will have peculiar properties which
  distinguishes a quark star from a neutron star
  \cite{usov98:a,usov01:c,page02:a,usov01:a}.

\item A very high-luminosity flare took place in the Large Magellanic Cloud
  (LMC), some 55~kpc away, on 5~March 1979.  Another giant flare was observed
  on 27 August 1998 from \sgrnineteenoo.  The inferred peak luminosities for
  both events is $\sim 10^7$ times the Eddington limit for a solar mass
  object, and the rise time is very much smaller than the time needed to drop
  $\sim 10^{25}$~g (about $10^{-8} \msun$) of normal material onto a neutron
  star.  Alcock \etal\ \cite{alcock86:a} suggested a detailed model for the
  5~March 1979 event burst which involves the particular properties of strange
  matter (see also \cite{usov01:a,horvath93:a}). The model assumes that a lump
  of strange matter of $\sim 10^{-8}\,\msun$ fell onto a rotating strange
  star. Since the lump is entirely made up of self-bound high-density matter,
  there would be only little tidal distortion of the lump, and so the duration
  of the impact can be very short, around $\sim 10^{-6}~\rms$, which would
  explain the observed rapid onset of the gamma ray flash. The light curves
  expected for such giant bursts \cite{usov98:a,usov01:c,usov01:b,cheng03:a}
  should posses characteristic features that are well within the capabilities
  of ESA's INTErnational Gamma-Ray Astrophysics Laboratory (INTEGRAL
  \cite{integral02:a}) launched by the European Space Agency in October of
  2002.

\end{itemize}

\section*{Acknowledgments}

This material is based upon work supported by the National Science Foundation
under Grant No. 0457329, and by the Research Corporation.

\printindex

\begin{thebibliography}{100}

\bibitem{hessels06:a}
J. W. T. Hessels, S. M. Ransom, I. H. Stairs, P. C. C. Freire, V. M. Kaspi, and
  F. Camilo, Science {\bf 311} (2006) 1901.

\bibitem{backer82:a}
D. C. Backer, S. R. Kulkarni, C. Heiles, M. M. Davis, and W. M. Goss, Nature
  {\bf 300} (1982) 615.

\bibitem{fruchter88:a}
A. S. Fruchter, D. R. Stinebring, and J. H. Taylor, Nature {\bf 334} (1988)
  237.

\bibitem{kaaret06:a}
P. Kaaret {\it et al.}, Astrophys.\ J.\ {\bf 657} (2007) L97.

\bibitem{villain06:a}
L. Villain, EAS Publ.\ Ser.\ 21 (2006) 335.

\bibitem{glen97:book}
N. K. Glendenning, {\it Compact Stars, Nuclear Physics, Particle Physics, and
  General Relativity}, 2nd ed.\ (Springer-Verlag, New York, 2000).

\bibitem{weber99:book}
F. Weber, {\it Pulsars as Astrophysical Laboratories for Nuclear and Particle
  Physics}, High Energy Physics, Cosmology and Gravitation Series (IOP
  Publishing, Bristol, Great Britain, 1999).

\bibitem{heiselberg00:a}
H. Heiselberg and V. Pandharipande, Ann.\ Rev.\ Nucl.\ Part.\ Sci.\ {\bf 50}
  (2000) 481.

\bibitem{lattimer01:a}
J. M. Lattimer and M. Prakash, Astrophys.\ J.\ {\bf 550} (2001) 426.

\bibitem{blaschke01:trento}
{\it Physics of Neutron Star Interiors}, ed.\ by D.\ Blaschke, N.\ K.\
  Glendenning, and A.\ Sedrakian, Lecture Notes in Physics {\bf 578}
  (Spring-Verlag, Berlin, 2001).

\bibitem{weber05:a}
F. Weber, Prog.\ Part.\ Nucl.\ Phys.\ {\bf 54} (2005) 193, ({\tt
  astro-ph/0407155}).

\bibitem{sedrakian06:a}
A. Sedrakian, Prog.\ Part.\ Nucl.\ Phys.\  {\bf 58} (2007) 168.

\bibitem{page06:review}
D. Page and S. Reddy, Ann.\ Rev.\ Nucl.\ Part.\ Sci. {\bf 56} (2006) 327.

\bibitem{blaschke06:a} 
D. Blaschke and H. Grigorian, 
{\it Unmasking neutron star interiors using cooling simulations}, to appear
in Prog. Part. Nucl. Phys., ({\tt arXiv:astro-ph/0612092}).

\bibitem{bodmer71:a}
A. R. Bodmer, Phys.\ Rev.\ D {\bf 4} (1971) 1601.

\bibitem{witten84:a}
E. Witten, Phys.\ Rev.\ D {\bf 30} (1984) 272.

\bibitem{terazawa89:a}
H. Terazawa, INS-Report-338 (INS, Univ.\ of Tokyo, 1979); J.\ Phys.\ Soc.\
  Japan, {\bf 58} (1989) 3555; {\bf 58} (1989) 4388; {\bf 59} (1990) 1199.

\bibitem{alcock86:a}
C. Alcock, E. Farhi, and A. V. Olinto, Astrophys.\ J.\ {\bf 310} (1986) 261.

\bibitem{alcock88:a}
C. Alcock and A. V. Olinto, Ann.\ Rev.\ Nucl.\ Part.\ Sci.\ {\bf 38} (1988)
  161.

\bibitem{madsen98:b}
J. Madsen, Lecture Notes in Physics {\bf 516} (1999) 162.

\bibitem{glen94:a}
N. K. Glendenning, Ch. Kettner, and F. Weber, Phys.\ Rev.\ Lett.\ {\bf 74}
  (1995) 3519.

\bibitem{rajagopal01:a}
K. Rajagopal and F. Wilczek, {\it The Condensed Matter Physics of QCD}, At the
  Frontier of Particle Physics / Handbook of QCD, ed.\ M.\ Shifman, (World
  Scientific) (2001).

\bibitem{alford01:a}
M. Alford, Ann.\ Rev.\ Nucl.\ Part.\ Sci.\ {\bf 51} (2001) 131.

\bibitem{alford98:a}
M. Alford, K. Rajagopal, and F. Wilczek, Phys.\ Lett.\ {\bf 422B} (1998) 247.

\bibitem{rapp98+99:a}
R. Rapp, T. Sch{\"{a}}fer, E. V. Shuryak, and M. Velkovsky, Phys.\ Rev.\ Lett.\
  {\bf 81} (1998) 53; Ann.\ Phys.\ {\bf 280} (2000) 35.

\bibitem{oppenheimer39}
J. R. Oppenheimer and G. M. Volkoff, Phys.\ Rev.\ {\bf 55} (1939) 374.

\bibitem{tolman39:a}
R. C. Tolman, Phys.\ Rev.\ {\bf 55} (1939) 364.

%\bibitem{hessels06:a}
%J. W. T. Hessels, S. M. Ransom, I. H. Stairs, P. C. C. Freire, V. M. Kaspi, and
%  F. Camilo, Science {\bf 311} (2006) 1901.

%\bibitem{backer82:a}
%D. C. Backer, S. R. Kulkarni, C. Heiles, M. M. Davis, and W. M. Goss, Nature
%  {\bf 300} (1982) 615.

\bibitem{taylor89:a}
J. H. Taylor and J. M. Weisberg, Astrophys.\ J.\ {\bf 345} (1989) 434.

\bibitem{thorsett99:a}
S. E. Thorsett and D. Chakrabarty, Astrophys.\ J.\ {\bf 512} (1999) 288.

\bibitem{nice05:b}
D. J. Nice, E. M. Splaver, I. H. Stairs, O. Loehmer, A. Jessner, M. Kramer, and
  J. M. Cordes, {\it A 2.1 solar mass pulsar measured by relativistic orbital
  decay}, ({\tt astro-ph/0508050}).

\bibitem{barret06:a}
D. Barret, J.-F. Olive, and M. C. Miller, {\it The coherence of kHz
  quasi-periodic oscillations in the X-rays from accreting neutron stars},
  ({\tt astro-ph/0605486}).

\bibitem{barziv01:a}
O. Barziv, L. Kaper, M. H. van Kerkwijk, J. H. Telting, and J. van Paradijs,
  Astron.\ {\&} Astrophys.\ {\bf 377} (2001) 925.

\bibitem{casares98:a}
J. Casares, P. A. Charles, and E. Kuulkers, Astrophys.\ J.\ {\bf 493} (1998)
  L39.

\bibitem{orosz99:a}
J. A. Orosz and E. Kuulkers, Mon.\ Not.\ R.\ Astron.\ Soc.\ {\bf 305} (1999)
  132.

\bibitem{clark02:a}
J. S. Clark, S. P. Goodwin, P. A. Crowther, L. Kaper, M. Fairbairn, N. Langer,
  and C. Brocksopp, Astron.\ {\&} Astrophys.\ {\bf 392} (2002) 909.

\bibitem{shahbaz04:a}
T. Shahbaz, J. Casares, C. A. Watson, P. A. Charles, R. I. Hynes, S. C. Shih,
  and D. Steeghs, Astrophys.\ J.\ {\bf 616} (2004) L123.

\bibitem{rhoades74:a}
C. E. Rhoades and R. Ruffini, Phys.\ Rev.\ Lett.\ {\bf 32} (1974) 324.

\bibitem{sabbadini77:a}
A. G. Sabbadini and J. B. Hartle, Ann.\ Phys.\ (N.Y.) {\bf 104} (1977) 95.

\bibitem{hartle78:a}
J. B. Hartle, Phys.\ Rep.\ {\bf 46} (1978) 201.

\bibitem{brown94:a}
G. E. Brown and H. A. Bethe, Astrophys.\ J.\ {\bf 423} (1994) 659.

\bibitem{myers95:a}
W. D. Myers and W. J. Swiatecki, Nucl.\ Phys.\ {\bf A601} (1996) 141.

\bibitem{strobel97:a}
K. Strobel, F. Weber, M. K. Weigel, and Ch. Schaab, Int.\ J.\ Mod.\ Phys.\ E
  {\bf 6}, No.\ 4 (1997) 669.

\bibitem{pandharipande79:a}
V. R. Pandharipande and R. B. Wiringa, Rev.\ Mod.\ Phys.\ {\bf 51} (1979) 821.

\bibitem{wiringa88:a}
R. B. Wiringa, V. Fiks, and A. Fabrocini, Phys.\ Rev.\ C {\bf 38} (1988) 1010.

\bibitem{akmal98:a}
A. Akmal, V. R. Pandharipande, and D. G. Ravenhall, Phys.\ Rev.\ C {\bf 58}
  (1998) 1804.

\bibitem{baldo99:BBG}
M. Baldo, G. F. Burgio, and H. J. Schulze, Phys.\ Rev.\  C {\bf 61} (2000)
055801.

\bibitem{baldo01:springer}
M. Baldo and F. Burgio, Lect. Notes Phys. {\bf 578} (2001) 1.

\bibitem{burgio02:a}
G. F. Burgio, M. Baldo, H.-J. Schulze, and P. K. Sahu, Phys.\ Rev.\ C {\bf 66} 
(2002) 025802.

\bibitem{lenske95:a}
H. Lenske and C. Fuchs, Phys.\ Lett.\ {\bf 345B} (1995) 355.

\bibitem{fuchs95:a}
C. Fuchs, H. Lenske, and H. H. Wolter, Phys.\ Rev.\ C {\bf 52} (1995) 3043.

\bibitem{typel99:a}
S. Typel and H. H. Wolter, Nucl.\ Phys.\ {\bf A656} (1999) 331.

\bibitem{hofmann01:a}
F. Hofmann, C. M. Keil, and H. Lenske, Phys.\ Rev.\ C {\bf 64} (2001) 034314.

\bibitem{niksic02:a}
T. Nik\v{s}i\'{c}, D. Vretenar, P.Finelli and P. Ring, Phys.\ Rev.\ C {\bf 66}
  (2002) 024306.

\bibitem{ban04:a}
S. F. Ban, J. Li, S. Q. Zhang, H. Y. Jia, and J. P. Sang, and J. Meng, Phys.\
  Rev.\ C {\bf 69} (2004) 045805.

\bibitem{buballa05:a}
M. Buballa, Phys.\ Rept.\ {\bf 407} (2005) 205.

\bibitem{blaschke05:a}
D. Blaschke, S. Fredriksson, H. Grigorian, A. M. {\"{O}}ztas and F. Sandin,
  Phys.\ Rev.\ D {\bf 72} (2005) 065020.

\bibitem{rischke05:a}
S. B. Ruster, V. Werth, M. Buballa, I. A. Shovkovy, and D. H. Rischke, Phys.\
  Rev.\ D {\bf 72} (2005) 034004.

\bibitem{abuki06:a}
H. Abuki and T. Kunihiro, Nucl.\ Phys.\ A {\bf 768} (2006) 118.

\bibitem{lawley06:a}
S. Lawley, W. Bentz, and A. W. Thomas, J.\ Phys.\ G: Nucl.\ Part.\ Phys.\ {\bf
  32} (2006) 667.

\bibitem{lawley06:b}
S. Lawley, W. Bentz, and A. W. Thomas, Phys.\ Lett.\ {\bf B632} (2006) 495.

\bibitem{wang05:a}
P. Wang, S. Lawley, D. B. Leinweber, A. W. Thomas, and A. G. Williams, Phys.\
  Rev.\ C {\bf 72} (2005) 045801.

\bibitem{glen92:crust}
N. K. Glendenning and F. Weber, Astrophys.\ J.\ {\bf 400} (1992) 647.

\bibitem{stejner05:a}
M. Stejner and J. Madsen, Phys.\ Rev.\ D {\bf 72} (2005) 123005.

\bibitem{glen85:b}
N. K. Glendenning, Astrophys.\ J.\ {\bf 293} (1985) 470.

\bibitem{huber98:a}
H. Huber, F. Weber, M. K. Weigel, and Ch. Schaab, Int.\ J.\ Mod.\ Phys.\ E {\bf
  7}, No.\ 3 (1998) 301.

\bibitem{prakash97:a}
M. Prakash, I. Bombaci, M. Prakash, P. J. Ellis, J. M. Lattimer, and R.
  Knorren, Phys.\ Rep.\ {\bf 280} (1997) 1.

\bibitem{lattimer91:a}
J. M. Lattimer, C. J. Pethick, M. Prakash, and P. Haensel, Phys.\ Rev.\ Lett.\
  {\bf 66} (1991) 2701.

\bibitem{prakash92:a}
M. Prakash, M. Prakash, J. M. Lattimer, and C. J. Pethick, Astrophys.\ J.\ {\bf
  390} (1992) L77.

\bibitem{haensel94:a}
P. Haensel and O. Yu. Gnedin, Astron.\ {\&} Astrophys. {\bf 290} (1994) 458.

\bibitem{schaab95:a}
Ch. Schaab, F. Weber, M. K. Weigel, and N. K. Glendenning, Nucl.\ Phys.\ {\bf
  A605} (1996) 531.

\bibitem{baym78:a}
G. Baym, {\it Neutron Stars and the Physics of Matter at High Density}, in:
  Nuclear Physics with Heavy Ions and Mesons, Vol.\ 2, Les Houches, Session
  XXX, ed.\ by R.\ Balian, M.\ Rho and G.\ Ripka (North-Holland, Amsterdam,
  1978) p.\ 745.

\bibitem{barshay73:a}
S. Barshay and G. E. Brown, Phys.\ Lett.\ {\bf 47B} (1973) 107.

\bibitem{brown88:a}
G. E. Brown, K. Kubodera, D. Page, and P. Pizzochero, Phys.\ Rev.\ D {\bf 37}
  (1988) 2042.

\bibitem{kaplan86:a}
D. B. Kaplan and A. E. Nelson, Phys.\ Lett.\ {\bf 175B} (1986) 57; {\it ibid.}
  Nucl.\ Phys. {\bf A479} (1988) 273.

\bibitem{brown87:a}
G. E. Brown, K. Kubodera, and M. Rho, Phys.\ Lett.\ {\bf 192B} (1987) 273.

\bibitem{lee95:a}
C.-H. Lee and M. Rho, {\it Kaon condensation in dense stellar matter}, Proc.\
  of the International Symposium on Strangness and Quark Matter, ed.\ by G.\
  Vassiliadis, A.\ Panagiotou, B.\ S.\ Kumar, and J.\ Madsen (World Scientific,
  Singapore, 1995) p.\ 283.

\bibitem{barth97:a}
R. Barth et al., Phys.\ Rev.\ Lett.\ {\bf 78} (1997) 4007.

\bibitem{senger01:a}
P. Senger, Nucl.\ Phys.\ {\bf A685} (2001) 312c.

\bibitem{sturm01:a}
C. Sturm et al., Phys.\ Rev.\ Lett.\ {\bf 86} (2001) 39.

\bibitem{devismes02:a}
A. Devismes, J.\ Phys.\ G: Nucl.\ Part.\ Phys.\ {\bf 28} (2002) 1591.

\bibitem{fuchs06:a}
Ch. Fuchs, Prog.\ Part.\ Nucl.\ Phys.\ {\bf 56} (2006) 1.

\bibitem{li97:a}
G. Q. Li, C.-H. Lee, and G. E. Brown, Nucl.\ Phys.\ {\bf A625} (1997) 372.

\bibitem{li97:b}
G. Q. Li, C.-H. Lee, and G. E. Brown, Phys.\ Rev.\ Lett.\ {\bf 79} (1997) 5214.

\bibitem{brown97:a}
G. E. Brown, Phys.\ Bl.\ {\bf 53} (1997) 671.

\bibitem{brown96:a}
G. E. Brown, {\it Supernova Explosions, Black Holes and Nucleon Stars}, in:
  Proceedings of the Nuclear Physics Conference -- INPC '95, ed.\ by S.\ Zuxun
  and X.\ Jincheng (World Scientific, Singapore, 1996) p.\ 623.

\bibitem{thorsson94:a}
V. Thorsson, M. Prakash, and J. M. Lattimer, Nucl.\ Phys.\ {\bf A572} (1994)
  693.

\bibitem{thielemann90:a}
F.-K. Thielemann, M.-A. Hashimoto, and K. Nomoto, Astrophys.\ J.\ {\bf 349}
  (1990) 222.

\bibitem{jaffe77:a}
R. L. Jaffe, Phys.\ Lett.\ {\bf 38} (1977) 195.

\bibitem{tamagaki91:a}
R. Tamagaki, Prog.\ Theor.\ Phys.\ {\bf 85} (1991) 321.

\bibitem{sakai97:a}
T. Sakai, J. Mori, A. J. Buchmann, K. Shimizu, and K. Yazaki, Nucl.\ Phys.\
  {\bf A625} (1997) 192.

\bibitem{glen98:a}
N. K. Glendenning and J. Schaffner-Bielich, Phys.\ Rev.\ C {\bf 58} (1998)
  1298.

\bibitem{faessler97:a}
A. Faessler, A. J. Buchmann, M. I. Krivoruchenko, and B. V. Martemyanov, Phys.\
  Lett.\ {\bf 391B} (1997) 255.

\bibitem{faessler97:b}
A. Faessler, A. J. Buchmann, and M. I. Krivoruchenko, Phys.\ Rev.\ C {\bf 56}
  (1997) 1576.

\bibitem{jaffe03:a}
R. Jaffe and F. Wilczek, Phys.\ Rev.\ Lett.\ {\bf 91} (2003) 232003.

\bibitem{jaffe05:a}
R. L. Jaffe, Phys.\ Rep.\ {\bf 409} (2005) 1; Nucl.\ Phys.\ Proc.\ Suppl.\ {\bf
  142} (2005) 343.

\bibitem{ivanenko65:a}
D. D. Ivanenko and D. F. Kurdgelaidze, Astrophys.\ {\bf 1} (1965) 251.

\bibitem{itoh70:a}
N. Itoh, Progr. Theor. Phys. {\bf 44} (1970) 291.

\bibitem{fritzsch73:a}
H. Fritzsch, M. Gell--Mann, and H. Leutwyler, Phys.\ Lett.\ {\bf 47B} (1973)
  365.

\bibitem{baym76:a}
G. Baym and S. Chin, Phys.\ Lett.\ {\bf 62B} (1976) 241.

\bibitem{keister76:a}
B. D. Keister and L. S. Kisslinger, Phys.\ Lett.\ {\bf 64B} (1976) 117.

\bibitem{chap77:a}
G. Chapline and M. Nauenberg, Phys.\ Rev.\ D {\bf 16} (1977) 450.

\bibitem{fech78:a}
W. B. Fechner and P. C. Joss, Nature {\bf 274} (1978) 347.

\bibitem{chap77:b}
G. Chapline and M. Nauenberg, Ann.\ New York Academy of Sci.\ {\bf 302} (1977)
  191.

\bibitem{glen91:pt}
N. K. Glendenning, Phys.\ Rev.\ D {\bf 46} (1992) 1274.

\bibitem{glen97:a}
N. K. Glendenning, S. Pei, and F. Weber, Phys.\ Rev.\ Lett.\ {\bf 79} (1997)
  1603.

\bibitem{chodos74:a}
A. Chodos, R. L. Jaffe, K. Johnson, C. B. Thorne, and V. F. Weisskopf, Phys.\
  Rev.\ D {\bf 9} (1974) 3471.

\bibitem{chodos74:b}
A. Chodos, R. L. Jaffe, K. Johnson, and C. B. Thorne, Phys.\ Rev.\ D {\bf 10}
  (1974) 2599.

\bibitem{weber06:a}
F. Weber, A. Torres i Cuadrat, A. Ho, and P. Rosenfield, ({\tt
  astro-ph/0602047}).

\bibitem{weber99:topr}
F. Weber, J.\ Phys.\ G: Nucl.\ Part.\ Phys.\ {\bf 25} (1999) R195.

\bibitem{glen00:b}
N. K. Glendenning and F. Weber, {\it Signal of Quark Deconfinement in
  Millisecond Pulsars and Reconfinement in Accreting X-ray Neutron Stars},
  Lecture Notes in Physics {\bf 578}, (Springer-Verlag, Berlin, 2001), p.\ 305.

\bibitem{alford03:a}
M. Alford, C. Kouvaris, and K. Rajagopal, Phys.\ Rev.\ Lett.\ {\bf 92} (2004)
  222001.

\bibitem{rajagopal01:b}
K. Rajagopal and F. Wilczek, Phys.\ Rev.\ Lett.\ {\bf 86} (2001) 3492.

\bibitem{bedaque01:a}
P. F. Bedaque and T. Sch{\"{a}}fer, Nucl.\ Phys.\ {\bf A697} (2002) 802.

\bibitem{kaplan02:a}
D. B. Kaplan and S. Reddy, Phys.\ Rev.\ D {\bf 65} (2002) 054042.

\bibitem{buballa02:a}
M. Buballa, J. Hosek and M. Oertel, Phys.\ Rev.\ Lett.\ {\bf 90} (2003) 182002.

\bibitem{schmitt04:a}
A. Schmitt, {\it Spin-one Color Superconductivity in Cold and Dense Quark
  Matter}, Ph.D.\ thesis, {\tt nucl-th/0405076}.

\bibitem{schaefer00:a}
T. Sch{\"{a}}fer, Phys.\ Rev.\ D {\bf 62} (2000) 094007.

\bibitem{alford00:a}
M. Alford, J. A. Bowers, and K. Rajagopal, Phys.\ Rev.\ D {\bf 63} (2001)
  074016.

\bibitem{bowers02:a}
J. A. Bowers and K. Rajagopal, Phys.\ Rev.\ D {\bf 66} (2002) 065002.

\bibitem{casalbuoni04:a}
R. Casalbuoni and G. Nardulli, Rev.\ Mod.\ Phys.\ {\bf 76} (2004) 263.

\bibitem{alford99:b}
M. Alford, K. Rajagopal, and F. Wilczek, Nucl. Phys. {\bf B537} (1999) 443.

\bibitem{son99:a}
D. T. Son, Phys.\ Rev.\ D {\bf D59} (1999) 094019.

\bibitem{alford03:b}
M. Alford and S. Reddy, Phys.\ Rev.\ D {\bf 67} (2003) 074024.

\bibitem{alford04:a}
M. Alford, J.\ Phys.\ G {\bf 30} (2004) S441.

\bibitem{rajagopal00:a}
K. Rajagopal, Acta Physica Polonica B {\bf 31} (2000) 3021.

\bibitem{alford00:b}
M. Alford, J. A. Bowers, and K. Rajagopal, J.\ Phys.\ G {\bf 27} (2001) 541.

\bibitem{blaschke99:a}
D. Blaschke, D. M. Sedrakian, and K. M. Shahabasyan, Astron.\ {\&} Astrophys.\
  {\bf 350} (1999) L47.

\bibitem{alford06:a}
M. Alford, K. Rajagopal, S. Reddy, and A. W. Steiner, Phys.\ Rev.\ D {\bf 73}
  (2006) 114016.

\bibitem{madsen88:a}
J. Madsen, Phys.\ Rev.\ Lett.\ {\bf 61} (1988) 2909.

\bibitem{mathews06:a}
G. J. Mathews, I.-S. Suh, B. O'Gorman, N. Q. Lan, W. Zech, K. Otsuki, and F.
  Weber, J.\ Phys.\ G: Nucl.\ Part.\ Phys.\ {\bf 32} (2006) 1.

\bibitem{kettner94:b}
Ch. Kettner, F. Weber, M. K. Weigel, and N. K. Glendenning, Phys.\ Rev.\ D {\bf
  51} (1995) 1440.

\bibitem{jaikumar05:a}
P. Jaikumar, S. Reddy, A. W. Steiner, Phys.\ Rev.\ Lett.\ {\bf 96} (2006)
  041101.

\bibitem{provencal98:a}
J. L. Provencal, H. L. Shipman, E. Hog, and P. Thejll, Astrophys.\ J.\ {\bf
  494} (1998) 759.

\bibitem{provencal02:a}
J. L. Provencal, H. L. Shipman, D. Koester, F. Wesemael, and P. Bergeron,
  Astrophys.\ J.\ {\bf 568} (2002) 324.

\bibitem{kepler00:a}
S. O. Kepler et al., Astrophys.\ J.\ {\bf 539} (2000) 379.

\bibitem{mathews04:a}
G. J. Mathews, B. O'Gorman, K. Otsuki, I. Suh, and F. Weber, Univ.\ of Notre
  Dame preprint (2003).

\bibitem{usov98:a}
V. V. Usov, Phys.\ Rev.\ Lett.\ {\bf 80} (1998) 230.

\bibitem{usov01:c}
V. V. Usov, Astrophys.\ J.\ {\bf 550} (2001) L179.

\bibitem{usov01:b}
V. V. Usov, Astrophys.\ J.\ {\bf 559} (2001) L137.

\bibitem{cheng03:a}
K. S. Cheng and T. Harko, Astrophys. J. {\bf 596} (2003) 451.

\bibitem{vogt03:a}
C. Vogt, R. Rapp, and R. Ouyed, Nucl.\ Phys.\ {\bf A735} (2004) 543.

\bibitem{weber06:iship}
F. Weber, M. Meixner, R. P. Negreiros, and M. Malheiro, {\it Ultra-Dense
  Neutron Star Matter, Strange Quark Stars, and the Nuclear Equation of State},
  ({\tt astro-ph/0606093}).

\bibitem{glen85:a}
N. K. Glendenning, Phys.\ Lett.\ {\bf 114B} (1982) 392; \\ N. K. Glendenning,
  Astrophys.\ J.\ {\bf 293} (1985) 470; \\ N. K. Glendenning, Z.\ Phys.\ A {\bf
  326} (1987) 57; \\ N. K. Glendenning, Z.\ Phys.\ A {\bf 327} (1987) 295.

\bibitem{lindblom01:proc}
L. Lindblom, {\it Neutron Star Pulsations and Instabilities}, in: Gravitational
  Waves: A Challenge to Theoretical Astrophysics, edited by V.\ Ferrari, J.\
  C.\ Miller, and L.\ Rezzolla, ICTP Lecture Notes Series, Vol. III, (ISBN
  92-95003-05-5, May 2001), ({\tt astro-ph/0101136}).

\bibitem{andersson98:a}
N. Andersson, Astrophys.\ J.\ {\bf 502} (1998) 708.

\bibitem{andersson01:a}
N. Andersson and K. Kokkotas, Int.\ J.\ Mod.\ Phys.\ {\bf D10} (2001) 381.

\bibitem{lindblom02:a}
L. Lindblom and B. Owen, Phys.\ Rev. D {\bf 65} (2002) 063006.

\bibitem{friedman98:a}
J. L. Friedman and S. M. Morsink, Astrophys.\ J.\ {\bf 502} (1998) 714.

\bibitem{madsen98:a}
J. Madsen, Phys.\ Rev.\ Lett.\ {\bf 81} (1998) 3311.

\bibitem{madsen00:b}
J. Madsen, Phys.\ Rev.\ Lett. {\bf 85} (2000) 10.

\bibitem{klis00:a}
M. van der Klis, Ann.\ Rev.\ Astron.\ Astrophys.\ {\bf 38} (2000) 717.

\bibitem{manuel04:a}
C. Manuel, A. Dobado, and F. J. Llanes-Estrada, {\it Shear Viscosity in a CFL
  Quark Star}, ({\tt hep-ph/0406058}).

\bibitem{usov04:a}
V. V. Usov, Phys.\ Rev.\ D {\bf 70} (2004) 067301.

\bibitem{ray03:a}
S. Ray, A. L. Esp\'{i}ndola, M. Malheiro. J. P. S. Lemos, and V. T. Zanchin,
  Phys.\ Rev.\ D {\bf 68} (2003) 084004.

\bibitem{bekenstein71:a}
J. D. Bekenstein, Phys. Rev. D {\bf 4} (1971) 2185.

\bibitem{ams01:homepage}
The AMS home page is {\tt http://ams.cern.ch}.

\bibitem{sandweiss04:a}
J. Sandweiss, J.\ Phys.\ G: Nucl.\ Part.\ Phys.\ {\bf 30} (2004) S51.

\bibitem{saito90:a}
T. Saito, Y. Hatano, Y. Fukuda, and H. Oda, Phys.\ Rev.\ Lett.\ {\bf 65} (1990)
  2094.

\bibitem{saito94:a}
T. Saito, {\it Test of the CRASH experiment counters with heavy ions}, Proc.\
  of the International Symposium on Strangeness and Quark Matter, ed.\ by G.\
  Vassiliadis, A.\ D.\ Panagiotou, B.\ S.\ Kumar, and J.\ Madsen (World
  Scientific, Singapore, 1995) p.\ 259.

\bibitem{ichimura93:a}
M. Ichimura et al., Nuovo Cimento A {\bf 36} (1993) 843.

\bibitem{ecco01:homepage}
Information about ECCO can be found at {\tt http://ultraman.berkeley.edu}.

\bibitem{shaulov96:a}
S. B. Shaulov, APH N.S., Heavy Ion Physics {\bf 4} (1996) 403.

\bibitem{rujula83:a}
A. De R{\'{u}}jula, S. L. Glashow, R. R. Wilson, and G. Charpak, Phys.\ Rep.\
  {\bf 99} (1983) 341.

\bibitem{miyamura95:a}
O. Miyamura, Proc.\ of the 24th International Cosmic Ray Conference, 1 (Rome,
  1995) p.\ 890.

\bibitem{lord95:a}
J. J. Lord and J. Iwai, Paper 515, presented at the International Conference on
  High Energy Physics, Dallas (1992); H. Wilczynski et al., Proceedings of the
  XXIV International Cosmic Ray Conference, HE Sessions, Rome (1995), Vol.\ 1,
  p.\ 1.

\bibitem{macro92:a}
MACRO Collaboration, Phys.\ Rev.\ Lett.\ {\bf 69} (1992) 1860.

\bibitem{ambrosio00:a}
M. Ambrosio et al., EPJ {\bf C13} (2000) 453.

\bibitem{ambrosio02:a}
M. Ambrosio et al., for the MACRO Collaboration, {\it Status Report of the
  MACRO Experiment for the year 2001 }, ({\tt hep-ex/0206027}).

\bibitem{giacomelli02:a}
G. Giacomelli, for the MACRO Collaboration, ({\tt hep-ex/0210021}).

\bibitem{lu04:a}
Z.-T. Lu, R. J. Holt, P. Mueller, T. P. O'Connor, J. P. Schiffer, and L.-B.
  Wang, {\it Searches for Stable Strangelets in Ordinary Matter: Overview and a
  Recent Example}, ({\tt nucl-ex/0402015}).

\bibitem{rujula84:a}
A. De R{\'{u}}jula and S. L. Glashow, Nature {\bf 312} (1984) 734.

\bibitem{price84:a}
P. B. Price, Phys.\ Rev.\ Lett.\ {\bf 52} (1984) 1265.

\bibitem{bruegger89:a}
M. Br{\"{u}}gger, K. L{\"{u}}tzenkirchen, S. Polikanov, G. Herrmann, M.
  Overbeck, N. Trautmann, A. Breskin, R. Chechik, Z. Fraenkel, and U.
  Smilansky, Nature {\bf 337} (1989) 434.

\bibitem{isaac98:a}
M. C. Perillo Isaac et al., Phys.\ Rev.\ Lett.\ {\bf 81} (1998) 2416; {\it
  ibid.} {\bf 82} (1999) 2220 (erratum).

\bibitem{thomas95:a}
J. Thomas and P. Jacobs, {\it A Guide to the High Energy Heavy Ion
  Experiments}, UCRL-ID-119181.

\bibitem{rusek96:a}
A. Rusek et al., (E886 collaboration), Phys.\ Rev.\ C {\bf 54} (1996) R15.

\bibitem{buren99:a}
G. Van Buren (E864 Collaboration), J.\ Phys.\ G: Nucl.\ Part.\ Phys.\ {\bf 25}
  (1999) 411.

\bibitem{belz96:a}
J. Belz et al., (BNL E888 collaboration), Phys.\ Rev.\ D {\bf 53} (1996) R3487.

\bibitem{belz96:b}
J. Belz et al., (BNL E888 collaboration), Phys.\ Rev.\ Lett.\ {\bf 76} (1996)
  3277.

\bibitem{dittus95:a}
F. Dittus et al. (NA52 collaboration), {\it First look at NA52 data on Pb--Pb
  interactions at 158 A~GeV/c}, International Conference on Strangeness in
  Hadronic Matter, ed.\ by J.\ Rafelski, AIP 340 (American Institute of
  Physics, New York, 1995) p.\ 24.

\bibitem{appelquist96:a}
G. Appelquist et al., Phys.\ Rev.\ Lett.\ {\bf 76} (1996) 3907.

\bibitem{ambrosini96:a}
G. Ambrosini et al., Nucl.\ Phys.\ {\bf A610} (1996) 306c.

\bibitem{klingenberg99:topr}
R. Klingenberg, J.\ Phys.\ G: Nucl.\ Part.\ Phys.\ {\bf 25} (1999) R273.

\bibitem{klahn06:a_short}
T. Klahn {\it et al.}, Phys.\ Rev.\ C {\bf 74} (2006) 035802.

\bibitem{lattimer05:a}
J. Lattimer and M. Prakash, Phys.\ Rev.\ Lett.\ {\bf 94} (2005) 111101.

\bibitem{alford06:b} 
M. Alford, D. Blaschke, A. Drago, T. Klahn, G. Pagliara, J.
Schaffner-Bielich, Nature {445} (2007) E7, ({\tt astro-ph/0606524}).

\bibitem{zdunik04:a}
J. L. Zdunik, P. Haensel, E. Gourgoulhon, and M. Bejger, Astron.\ {\&}
  Astrophys.\ {\bf 416} (2004) 1013.

\bibitem{chubarian00:a}
E. Chubarian, H. Grigorian, G. Poghosyan, and D. Blaschke, Astron.\ {\&}
  Astrophys.\ {\bf 357} (2000) 968.

\bibitem{page02:a}
D. Page and V. V. Usov, Phys.\ Rev.\ Lett.\ {\bf 89} (2002) 131101.

\bibitem{usov01:a}
V. V. Usov, Phys.\ Rev.\ Lett.\ {\bf 87} (2001) 021101.

\bibitem{horvath93:a}
J. E. Horvath, H. Vucetich, and O. G. Benvenuto, Mon.\ Not.\ R.\ Astron.\ Soc.\
  {\bf 262} (1993) 506.

\bibitem{integral02:a}
See, for instance, {\tt astro-ph/0207527}.

\end{thebibliography}
\end{document}